\documentclass[pra, twocolumn, superscriptaddress, letterpaper, showkeys]{revtex4}

\usepackage{graphicx}
\usepackage{amsfonts}
\usepackage{amsmath}	% Matrix notation
\usepackage{epstopdf}
\usepackage{appendix}
\usepackage{hyperref}
\usepackage{enumerate}
\usepackage{verbatim}

\begin{document}

% MACROS
%%%%%%%%%%%%%%%%%%%%%%%%%%%%%%%%%%%%%%%%%%%%%%%%%%%%%%%%%%%%%%%%%%%%%

\newcommand{\beq}{\begin{equation}}
\newcommand{\eeq}{\end{equation}}
\newcommand{\beqa}{\begin{eqnarray}}
\newcommand{\eeqa}{\end{eqnarray}}
\newcommand{\vecmu}{\vec{\mu}}
\newcommand{\vectau}{\vec{\tau}}
\newcommand{\lf}{\hfil \break \break}
\newcommand{\Ahat}{\hat{A}}
\newcommand{\Adag}{\hat{A}^{\dagger}}
\newcommand{\ahat}{\hat{a}}
\newcommand{\adag}{\hat{a}^{\dagger}}
\newcommand{\Bhattilde}{\tilde{\hat{B}}}
\newcommand{\Bdagtilde}{\tilde{\hat{B}}^{\dagger}}
\newcommand{\Bhat}{\hat{B}}
\newcommand{\Bdag}{\hat{B}^{\dagger}}
\newcommand{\chat}{\hat{c}}
\newcommand{\cdag}{\hat{c}^{\dagger}}
\newcommand{\Ohat}{\hat{O}}
\newcommand{\Odag}{\hat{O}^{\dagger}}
\newcommand{\Uhat}{\hat{U}}
\newcommand{\Udag}{\hat{U}^{\dagger}}
\newcommand{\Xihat}{\hat{\Xi}}
\newcommand{\Xidag}{\hat{\Xi}^{\dagger}}
\newcommand{\Pihat}{\hat{\Pi}}
\newcommand{\nsat}{n_{\mbox{\scriptsize s}}}

\newcommand{\ADis}{\hat{A}^{\mbox{\scriptsize (Dis.)}}}
\newcommand{\AHad}{\hat{A}^{\mbox{\scriptsize (Had.)}}}
\newcommand{\PiAPD}{\hat{\Pi}^{\mbox{\scriptsize (APD)}}}
\newcommand{\PiHD}{\hat{\Pi}^{\mbox{\scriptsize (HD)}}}
\newcommand{\PiPNRD}{\hat{\Pi}^{\mbox{\scriptsize (PNRD)}}}

\newcommand{\rhohat}{\hat{\rho}}
\newcommand{\shat}{\hat{\sigma}}
\newcommand{\ket}[1]{\mbox{$|#1\rangle$}}
\newcommand{\bra}[1]{\mbox{$\langle#1|$}}
\newcommand{\ketbra}[2]{\mbox{$|#1\rangle \langle#2|$}}
\newcommand{\braket}[2]{\mbox{$\langle#1|#2\rangle$}}
\newcommand{\bracket}[3]{\mbox{$\langle#1|#2|#3\rangle$}}
\newcommand{\mat}[1]{\overline{\overline{#1}}}
\newcommand{\hak}[1]{\left[ #1 \right]}
\newcommand{\vin}[1]{\langle #1 \rangle}
\newcommand{\abs}[1]{\left| #1 \right|}
\newcommand{\tes}[1]{\left( #1 \right)}
\newcommand{\braces}[1]{\left\{ #1 \right\}}
\newcommand{\sub}[1]{{\mbox{\scriptsize #1}}}
\newcommand{\com}[1]{\textcolor{red}{[\textit{#1}]}}
\newcommand{\novel}[1]{\textcolor{blue}{#1}}
\newcommand{\id}{\hat{\mathbb{I}}}
\newcommand{\Khat}{\hat{K}}
\newcommand{\sop}{\hat{\$}}
\newcommand{\argmax}{\operatornamewithlimits{argmax}}

% TITLE AND ABSTRACT
%%%%%%%%%%%%%%%%%%%%%%%%%%%%%%%%%%%%%%%%%%%%%%%%%%%%%%%%%%%%%%%%%%%%%

\title{Algebraic and algorithmic frameworks for optimized quantum measurements}

\author{Amine Laghaout}
\author{Ulrik L. Andersen}
\affiliation{Department of Physics, Technical University of Denmark, Building 309, 2800 Lyngby, Denmark}

\date{\today}

\begin{abstract}
Von Neumann projections are the main operations by which information can be extracted from the quantum to the classical realm. They are however static processes that do not adapt to the states they measure. Advances in the field of adaptive measurement have shown that this limitation can be overcome by ``wrapping'' the von Neumann projectors in a higher-dimensional circuit which exploits the interplay between measurement outcomes and measurement settings. Unfortunately, the design of adaptive measurement has often been \textit{ad hoc} and setup-specific. We shall here develop a unified framework for designing optimized measurements. Our approach is two-fold: The first is algebraic and formulates the problem of measurement as a simple matrix diagonalization problem. The second is algorithmic and models the optimal interaction between measurement outcomes and measurement settings as a cascaded network of conditional probabilities. Finally, we demonstrate that several figures of merit, such as Bell factors, can be improved by optimized measurements. This leads us to the promising observation that measurement detectors which---taken individually---have a low quantum efficiency can be be arranged into circuits where, collectively, the limitations of inefficiency are compensated for.
\end{abstract}


\maketitle

\tableofcontents		

%%%%%%%%%%%%%%%%%%%%%%%%%%%%%%%%%%%%%%%%%%%%%%%%%%%%%%%%%%%%%%%%%%%%%
\section{Introduction}
\label{sec:Introduction}

Even within the physics community, the idea of measurement all too often evokes specific laboratory devices, such as photon counters or homodyne detectors, to name a couple of examples from quantum optics. In other words, we are accustomed to reducing measurements to algebraic projections which are \textit{static} in Hilbert space. For instance, the ``field of view'' of a photon counter is immutably constrained along the diagonal of the Fock Hilbert space and cannot be redirected to peek at the off-diagonal terms which conceal potentially valuable phase information. This type of basic measurement is called a direct measurement or, alternatively, a von Neumann projection \cite{Ivanovic1988}.

Over the past few decades, several measurement schemes have been independently developed which have successfully overcome the limitation of direct measurements. These schemes, variously referred to as quantum receivers \cite{Kennedy1972, Dolinar1973}, quantum filtering measurements \cite{Belavkin1983, Carmichael1993, Bouten2009, Somaraju2013}, or adaptive measurements \cite{Wiseman1995, Wiseman1997, Wiseman1997a, Armen2002, Wiseman2009}, have demonstrated that passive detection devices can be augmented by multi-modal quantum circuits so as to gain optimal insight into the states under scrutiny. Although they differ in their motives, these advances in quantum measurement all have in common that they incorporate von Neumann projections into larger setups involving ancillary resources, controllable unitary operations, and (most often) some Bayesian logic governing feedback loops between the detection outcomes and the unitary operations. We shall interchangeably refer to these collective techniques as generalized---or adaptive---measurements. 

Although the theory of generalized measurements is well established, notably through the work of Neumark \cite{Neumark1940} and Kraus \cite{Kraus1983}, it remains under-exploited in the \textit{design} of experiments. Indeed, most of the experimental advances cited above were arrived at in an \textit{ad hoc} fashion where heuristic approaches and setup-dependent models overlooked the bigger picture offered by Neumark's theorem. In all cases, the overarching goal can be stated as follows: Given the limited toolbox of detection devices and unitary operations that is readily available in the laboratory, how can one design---in a systematic way---a quantum measurement circuit that optimizes the relevance and accuracy of the acquired data? 

In the present article, we will show that a structured solution to this problem can be obtained on two fronts. The first one is algebraic: In Sec. \ref{sec:Algebraic}, we develop an intuition for Neumark's theorem which captures the essence of adaptive measurements. This will lead us to argue that the stochastic mindset which has so far dominated the theory of adaptive measurements \cite{Wiseman1993, Goetsch1994, Brun2001} is at the outset underpinned by a deterministic, albeit non-trivial, algebraic problem. We will show that this algebraic formulation  consists of finding a similarity transformation from a multi-mode sequence of direct measurements to an optimal positive-operator valued measure (POVM) in higher dimensions. The second aspect we shall bring forth is a computational, or algorithmic, one. This is what we treat in Sec. \ref{sec:Algorithmic} where we represent generalized measurements as Bayesian networks which, when laid out in a optimum way, can mimic the statistics of the aforementioned optimal POVM. Even if this approach does include conditional probabilities, its formulation is rather straightforward and does not resort to the stochastic machinery of quantum trajectories or master equations. Section \ref{sec:Simulations} discusses in detail the various figures of merit by which the efficiency of generalized measurements can be assessed. A selection of numerical simulations will be presented to illustrate the various trends of these figures of merit. It will then become apparent that, for any given detection device, what we usually think of as the limitations of quantum efficiency are really those of a direct measurement setup. However, if several such inefficient devices are judiciously assembled into a larger measurement circuit, the collective quantum efficiency is improved even if the building blocks, taken individually, are inefficient.

Before proceeding, let us clarify what is meant by the optimization of measurements. In quantum information, measurement is not only an end in itself, but can also be a means of computation and state preparation. Different figures of merit can therefore be subjected to optimization such as distinguishability measures \cite{Fuchs1996}, discrimination errors, Bell factors, state fidelity, etc. We shall introduce some of these figures of merit with a particular focus on the \textit{single shot discrimination} of quantum states. Indeed, we argue that the complete characterization of a state is really just a generalized discrimination problem where the set of possible states to be distinguished is infinite. Hence, any measurement is intrinsically a comparative operation that presupposes a pool of candidate states.

%%%%%%%%%%%%%%%%%%%%%%%%%%%%%%%%%%%%%%%%%%%%%%%%%%%%%%%%%%%%%%%%%%%%%
\section{Adaptive measurement as an algebraic problem}
\label{sec:Algebraic}

\subsection{An intuition for generalized measurements}

Let us first develop a basic intuition for generalized measurements before formally presenting  the problem of quantum state discrimination. Assume we are interested in discriminating a square from a circle which are drawn on a two-dimensional space such as a sheet of paper. From within the paper, both figures will appear as straight lines; their discrimination will therefore be impossible. However, if we step up to the third dimension, we will immediately be able to distinguish them even if their projection onto the detector---our eye---remains effectively two-dimensional. This is the essence of Neumark's theorem: By rising to higher dimensions in Hilbert space, we have the potential to recover information which was otherwise traced out or ``de-cohered'' in the reduced space containing the system of interest \cite{Bergou2010a, Bergou2010LectureNotes}. This said, it does not suffice to go up to higher dimensions to implement an efficient generalized measurement. If instead of comparing a circle and a square, we intend to compare an isoceles trapezoid and a square, not only will we have to include a third dimension, but  the \textit{vantage point} (in this case the Euclidean angle) from that third dimension will also have to be chosen carefully or else both figures may again be indistinguishable due to coinciding perspectives. This is especially crucial if we are constrained to a limited number of vantage points while trying to maximize the information gain about the measured objects. One could go about this problem in a stochastic way. E.g., we could start from one random vantage point and then, depending on what we ``see'', move in one direction or another so as to gradually increase the confidence in the discrimination. (This is the heuristic behind the Dolinar and Kennedy receivers \cite{Kennedy1972, Dolinar1973} as well as most of the subsequent schemes of adaptive measurement.) However, it is clear that the very nature of the problem is deterministic and the sequence of optimal vantage points can in principle be solved for exactly based solely on the (Hilbert) geometry of the states at play.

To the admittedly naive analogy above, one should add the extra complication that in quantum mechanics, the very act of measurement reshapes the geometrical figures (i.e., states) under observation---as dictated by the uncertainty principle. This fact introduces a uniquely quantum twist to our story: The optimal vantage points are not only based on the Hilbert space configuration of the objects to be distinguished, they are also interdependent and chronologically ordered. This will become clearer as we move on to a formal statement of the problem. 

\subsection{Adaptation as a similarity transformation}

Consider a pool of $C$ candidate states $\rhohat_{c}^{(0)} \in \mathcal{H}^{(0)}$, where $c \in \braces{1, \cdots, C}$, which are contained in a Hilbert space $\mathcal{H}^{(0)}$. No assumption is made as to purity or the mutual orthogonality of these states, except that they are normalized and that no two of them are exactly identical. Without loss of generality, we can assign to each state $c$ a prior probability $p_c^{(0)}$ that it be retrieved from the pool. Physically, these probabilities could represent the classical rate of incidence of the states onto the measurement apparatus. For completeness, we have
\beq
\sum\limits_{c=1}^{C} p_c^{(0)} = 1.
\eeq 

Furthermore, assume that there exists $M$ possible outcomes $\mu \in \braces{1, \cdots, M}$ at the end of the measurement \footnote{Throughout this article, we consider the measurement outcomes to be discrete. The generalization to continuous outcomes is however straightforward and does not detract from our argument.}. The prior probabilities $p_c^{(0)}$ will be redistributed among the different outcomes $\mu$, thereby creating probability distribution functions $p_c^{(1)}(\mu)$ such that
\beq
p_c^{(0)} = \sum\limits_{\mu=1}^{M} p_c^{(1)}(\mu).
\eeq
The parenthesized superscript indicates whether the probabilities pertain before (0) or after (1) the measurement. 

Ideally, we want the measurement operation to be a injective map \cite{Garnier2009} from the set of candidate quantum states $\braces{c}$ to that of the classical readouts $\braces{\mu}$. In other words, perfectly unambiguous discrimination is only possible if each readout $\mu$ is mapped by at most one candidate state $c$ while each candidate state $c$ maps to at least one readout $\mu$. (Such a non-overlapping, one-to-many mapping incidentally requires that $M \ge C$.) In general, however, this ideal condition is not likely to be met: The same outcome $\mu$ may be mapped by more than one state $c$ with varying probabilities, thereby introducing ambiguity in the discrimination. We are then faced with an optimization problem where the goal is to determine with highest confidence the classical identity $c$ of the unknown state. Since quantum states only manifest themselves to us through the probability distributions they cast on the measurement detectors, we shall claim that any two states are optimally discriminated if their probability distributions $p_c^{(1)}(\mu)$ are as dissimilar as possible. A more rigorous definition of ``dissimilarity'' will be provided in Sec. \ref{sec:Simulations} as one of the potential figures of merit $\mathcal{F}$ relevant to quantum measurements. For now, let us denote by $\Ohat_{\mu} \in \mathcal{H}^{(0)}$ the ideal POVM which produces these maximally dissimilar probability distributions or, more specifically, which maximizes $\mathcal{F}$, i.e.
\beq
\Ohat_{\mu} = \argmax_{\Ohat_{\mu}^{'}\in\mathcal{H}^{(0)}}{\mathcal{F}}.
\eeq
In quantum metrology, for example, $\Ohat_{\mu}$ could be any POVM which reaches the Heisenberg limit. (Note that any such ideal POVMs are guaranteed to exist as demonstrated in Refs. \cite{Reck1994, vanLoock2004}.)

Recall that Born's rule reads
\beq
p_c^{(1)}(\mu) = p_c^{(0)} \cdot \mbox{Tr}\braces{\Ohat_{\mu}\rhohat_{c} \Odag_{\mu}},
\label{eq:OhatBorn}
\eeq
where we have propagated the prior probability $p_c^{(0)}$, and that $\Ohat_{\mu}$ satisfies completeness
\beq
\sum\limits_{\mu=1}^{M} \Ohat_{\mu} = \hat{\mathbb{I}}.
\label{eq:OhatCompleteness}
\eeq

The measurement schemes we will devise shall strive to mimic the ideal POVM $\Ohat_{\mu}$, or at least reproduce the probability distributions it generates with as much fidelity as possible. We shall see see how generalized measurements perform better at this task than direct measurements. Let us first introduce the latter with a very generic notation that we shall maintain throughout the rest of this article.

\subsubsection{Direct measurements}
\label{sec:DirectMeasurements}

\begin{figure}[h]
  \includegraphics[width=0.45\columnwidth]{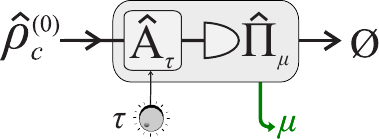}
  \caption{Generic representation of a direct measurement. The state is rotated by a unitary operation $\Ahat_{\tau}$ parametrized by $\tau$ and then collapsed by a projective measurement $\Pihat_{\mu}$ corresponding to one of $M$ possible readouts $\mu$. The empty set symbol $\emptyset$ indicates that the input state is irreversibly ``consumed'' by the end of the measurement at which point no further information can be retrieved.}
	\label{fig:vonNeumann}
\end{figure}

A direct measurement is represented generically in Fig. \ref{fig:vonNeumann}. The unknown candidate state undergoes a unitary operation $\Ahat_{\tau} \in \mathcal{H}^{(0)}$ which can be tuned by a  parameter (or set of parameters) $\tau$. This parameter could be any degree of freedom available to us in the laboratory, such as a coherent displacement, a squeezing factor, or the rotation angle induced by a set of wave plates. It is by tuning $\tau$ that we can search for the optimal vantage points discussed earlier. After this, the transformed state is collapsed by one of $M$ possible von Neumann projectors $\Pihat_{\mu} \in \mathcal{H}^{(0)}$ to each of which corresponds a classical output $\mu \in \braces{1, \cdots, M}$. If we assume that the detection device never fails to produce an output, the projectors should add up to unity
\beq
\sum\limits_{\mu=1}^{M} \Pihat_{\mu} = \hat{\mathbb{I}}.
\eeq
Inconclusive outcomes or failures in detection can be accounted for by allocating a fictitious outcome $\mu_{\mbox{\scriptsize fail}}$ among the $M$ outputs.

The probability that a state $c$ from the pool of candidate states triggers a readout $\mu$ is given by
\beq
p_c^{(1)}(\mu) = p_c^{(0)} \cdot p_{\mu\mid c}^{(1)},
\label{eq:DirectPropagation}
\eeq
where $p_{\mu\mid c}^{(1)}$ is the probability that a certain outcome $\mu$ is obtained given that state $c$ was incident (i.e., notwithstanding its prior probability). This conditional probability is given by
\beq
p_{\mu\mid c}^{(1)} = \mbox{Tr}\braces{\tes{\Adag_{\tau}\Pihat_{\mu} \Ahat_{\tau}}\rhohat_{c}^{(0)}}.
\label{eq:DirectBorn}
\eeq

Now that we have an expression for the probability distributions projected by the candidate states on the spectrum of outcomes, there remains to insure that these distributions are maximally discriminated. This can be achieved by tuning the controllable parameter $\tau$ such that the combination of $\Ahat_{\tau}$ and $\Pihat_{\mu}$ mimics as much as possible the statistics produced by the ideal POVM $\Ohat$. By comparing Eq. (\ref{eq:DirectBorn}) with Eq. (\ref{eq:OhatBorn}), we therefore need to solve for the optimal $\tau_{o}$ which best approximates a pseudo-similarity transformation from the von Neumann projector $\sqrt{\Pihat_{\mu}}$ to the ideal POVM $\Ohat_{\mu}$,
\beq
\sqrt{\Adag_{\tau_{o}} \Pihat_{\mu} \Ahat_{\tau_{o}}} \approx \Ohat_{\mu}, \, \forall \mu.
\label{eq:SimilarityDirect}
\eeq

Conceptually, what this transformation does in Hilbert space is to align the set of candidate states $\rhohat_{c}^{(0)}$ with the von Neumann projectors $\Pihat_{\mu}$. The unitary $\Ahat_{\tau}$ thus serves to present the states into a more revealing configuration in Hilbert space. In practice, however, because of the limited leeway offered by $\tau$, a strict equality in Eq. (\ref{eq:SimilarityDirect}) will be unlikely. Therefore, we may as well give a more operational statement of the problem whereby we search for the argument of the maximum for the figure of merit
\beq
\tau_{o} = \argmax_{\tau\in\mathcal{T}}{\mathcal{F}},
\eeq
where $\mathcal{T}$ is the parameter space, e.g., $[-\pi, \pi]$ for polarization rotations, or $\mathbb{C}$ for coherent state translations.

\subsubsection{Generalized measurements}

\begin{figure*}[ht]
  \centering\includegraphics[width=1.05\columnwidth]{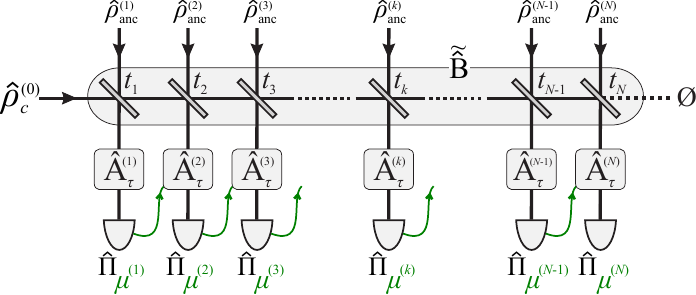}
  \caption{A generalized measurement is represented schematically as the multi-mode augmentation of a direct measurement. Whereas a direct measurement only spans the modes populated by the candidate states, a generalized measurement involves states and operations in ancillary modes, which---when chosen appropriately---provide a higher-dimensional perspective on the measured state. Here, we show the candidate state in the zeroth mode (the horizontal quantum channel) being coupled with known ancillary states in the modes 1 to $N$ (the vertical quantum channels). Taken individually, each ancillary mode then undergoes the same direct measurement process as described in Fig. \ref{fig:vonNeumann}. In contrast to the direct measurement, the generalized setup provides a finer-grained projection space for the probability distributions function with $M^N$ possible outcomes (as opposed to only $M$). Similarly, the generalized setup provides not just one, but $N$ degrees of freedom for tuning the unitary parameters $\tau^{(k)}$.}
	\label{fig:Setup}
\end{figure*}

We have just seen how a von Neumann projector can be amended with a unitary operation to improve the overall measurement efficacy. The flexibility afforded by the parameter $\tau$ can be used to  somewhat adjust the orientation of the quantum states with respect to the projection operator. This leeway is nonetheless constrained by the very nature of the unitary operations $\Ahat_{\tau}$ as well as their availability in our laboratory toolbox. Although it is always possible conceive of a better unitary $\Ahat_{\tau}^{'}$ \cite{Reck1994, vanLoock2004}, it may not exist physically or may simply be too demanding to engineer. This is where Neumark's theorem comes in. By rising to higher dimensions in Hilbert space, the measurement setup can be made even more flexible---i.e., adaptive---while still exploiting the same available building blocks of $\Ahat_{\tau}$ and $\Pihat_{\mu}$. This is achieved by coupling the unknown candidate state $\rhohat_{c}^{(0)}$ with $N$ known ancillary states $\rhohat_{\mbox{\scriptsize anc}}^{(k)} \in \mathcal{H}^{(k)}$, where $k \in \braces{1, \cdots, N}$. The parenthesized superscript labels the quantum modes: The zeroth mode is occupied by the input and all the ancillaries span modes 1 to $N$. The coupling of all $N+1$ modes, i.e., the ancillaries plus the unknown input, could be achieved by a beam splitting operation $\Bhattilde \in \bigotimes_{k=0}^{N} \, \mathcal{H}^{(k)}$. At each output of the beam splitters, one then grafts the same direct measurement described in Sec. \ref{sec:DirectMeasurements}. This multiplexed arrangement (Fig. \ref{fig:Setup}) of direct measurements will produce $N$ classical readouts---one from each mode---which we shall bundle into an array of length $N$
\beq
\vecmu_{l} = \hak{\mu^{(1)}, \cdots, \mu^{(k)}, \cdots, \mu^{(N)}},
\label{eq:vecnu}
\eeq
where $l \in \braces{1, \cdots, M^N}$ uniquely identifies one set of outcomes among all the possible combinations. 

What we shall consider from now on is therefore the probability distribution mapped by the candidate states $c$ onto the outcomes $\vecmu_{l}$. Just as in Eq. (\ref{eq:DirectPropagation}), these probability distributions are given by 
\beq
p_c^{(N)}(\vecmu_{l}) = p_c^{(0)} \cdot p_{\vecmu_{l} \mid c}^{(N)},
\label{eq:GeneralizedPropagation}
\eeq
Here again, the parenthesized superscript over the probabilities label the number of completed measurements: (0) indicates prior probabilities whereas ($N$) indicates that all $N$ measurements have been completed. (These labels should not be confused with the somewhat related superscripts over the density matrices of the ancillary states and their Hilbert spaces: Those indicate the quantum modes.)

The conditional probability that a certain outcome sequence $\vecmu_{l}$ is triggered by a state $c$ is given by
\begin{widetext}
\beqa
p_{\vecmu_{l}\mid c}^{(N)} & = & \mbox{Tr}\braces{\Bdagtilde \hak{\hat{\mathbb{I}} \otimes \bigotimes_{k=1}^{N} \Adag_{\tau^{(k)}} \Pihat_{\mu^{(k)}} \Ahat_{\tau^{(k)}}} \Bhattilde \tes{\rhohat_c^{(0)} \otimes \hak{\bigotimes_{k=1}^{N} \rhohat_{\mbox{\scriptsize anc}}^{(k)}}}} \nonumber\\
& = & \mbox{Tr}\braces{\Bdagtilde \hak{\hat{\mathbb{I}} \otimes \bigotimes_{k=1}^{N} \Adag_{\tau^{(k)}}} \hak{\hat{\mathbb{I}} \otimes \bigotimes_{k=1}^{N} \Pihat_{\mu^{(k)}}} \hak{\hat{\mathbb{I}} \otimes \bigotimes_{k=1}^{N} \Ahat_{\tau^{(k)}}} \Bhattilde \tes{\rhohat_c^{(0)} \otimes  \hak{\bigotimes_{k=1}^{N} \rhohat_{\mbox{\scriptsize anc}}^{(k)}}}} \nonumber\\
p_{\vecmu_{l}\mid c}^{(N)} & = & \mbox{Tr}\braces{\tes{\Uhat_{\vectau} \tilde{\Pihat}_{\vecmu} \Udag_{\vectau}} \tilde{\rhohat}_{c}^{(0)}},
\label{eq:GeneralizedBorn}
\eeqa
\end{widetext}
where we have assembled the coupling operator $\Bhattilde$ and the operations $\Ahat_{\tau^{(k)}}$ into one big unitary
\beq
\Uhat_{\vectau} = \Bdagtilde \hak{\hat{\mathbb{I}} \otimes \bigotimes_{k=1}^{N} \Ahat_{\tau^{(k)}}}.
\eeq
The $N$-dimensional array $\vectau \in \mathcal{T}^{\otimes N}$ represents a combination of parameter settings for the unitary operations at each mode
\beq
\vectau = \hak{\tau^{(1)}, \cdots, \tau^{(k)}, \cdots, \tau^{(N)}}.
\label{eq:vectau}
\eeq
Similarly, we have grouped the von Neumann projections and the input states into single matrices, indicated by a tilde, and spanning all $N+1$ modes:
\beqa
\tilde{\Pihat}_{\vecmu} & = & \hat{\mathbb{I}} \otimes \bigotimes_{k=1}^{N} \Pihat_{\mu^{(k)}}, \\
\tilde{\rhohat}_{c}^{(0)} & = & \rhohat_c^{(0)} \otimes \hak{\bigotimes_{k=1}^{N} \rhohat_{\mbox{\scriptsize anc}}^{(k)}}.
\eeqa
Note that since the ancillary states $\rhohat_{\mbox{\scriptsize anc}}^{(k)}$ are known and initially independent of the candidate states $\rhohat_c^{(0)}$, the information content of the augmented state $\tilde{\rhohat}_{c}^{(0)}$ is exactly the same as that of $\rhohat_c^{(0)}$.

If we draw the parallel between Eq. (\ref{eq:GeneralizedBorn}) and its direct measurement analog,  Eq. (\ref{eq:DirectBorn}), we see that we can again tweak the parameters $\vectau$ to approximate a pseudo-similarity transformation akin to that of Eq. (\ref{eq:SimilarityDirect})
\beq
\sqrt{\Udag_{\vectau_{o}} \tilde{\Pihat}_{\vecmu_{l}} \Uhat_{\vectau_{o}}} \approx \tilde{\Ohat}_{\vecmu_{l}}, \, \forall\vecmu_{l},
\label{eq:SimilarityGeneralized}
\eeq
where $\tilde{\Ohat}_{\vecmu_{l}}$ is the multi-mode counterpart of the ideal POVM $\Ohat_{\mu}$ conceived of in Eq. (\ref{eq:OhatBorn}). Alternatively, the parameter setting $\vectau_{o}$ which maximizes the measurement figure of merit can be found by optimization such that
\beq
\vectau_{o} = \argmax_{\vectau\in\mathcal{T}^{\otimes N}}{\mathcal{F}}.
\label{eq:ArgmaxGeneralized}
\eeq

We can already see that a generalized measurement offers a two-fold advantage over its direct  counterpart. The first is that the cardinality of the projection space, i.e., the set of classical outcomes, increases from $M$ to $M^N$, thereby opening up the possibility of discriminating more states than would be possible if $C > M$. Moreover, the increased range of classical readouts allows for a crisper resolution of the probability distributions. This can in principle be valuable in reducing their overlap, and therefore in reducing the ambiguity of the discrimination. (Recall that perfect discrimination requires an injective mapping from the set of candidate states $c$ to that of outcomes $\mu$.) The second advantage of generalized measurements is that they provide us with, not one, but $N$ ``tuning knobs'' $\vectau$. If we add to this the choice of $N$ ancillary states, it becomes clear that the generalized setup offers much more leeway to prepare the candidate states before they are irreversibly collapsed by the von Neumann projections.

\subsection{Adaptation as a dynamic problem}

So far, we have presented the optimization of measurement as a deterministic, one-off calculation based on what we know about the Hilbert space geometry of the input states and the projectors. This culminated with two algebraic formulations of the problem, namely Eq. (\ref{eq:SimilarityGeneralized}) and Eq. (\ref{eq:ArgmaxGeneralized}). One could be content with this understanding of measurement optimization as the search for an optimal---but static---vantage point $\vectau_{o}$ \footnote{We loosely refer to $\vectau_{o}$ as a ``vantage point''. Rigorously speaking, what is meant is that the unitary operations parametrized by $\vectau_{o}$ rotate the von Neumann projector in Hilbert space so as to achieve an optimal perspective on the candidate states.}. However, there exists yet a third and more crucial difference between direct and generalized measurements. Whereas wave function collapse occurs at once in the former, it has the potential to happen gradually in the latter. With each partial collapse up to mode $k$, the pool of candidate states is reshaped; hence, the optimal vantage points $\tau_{o}^{(k')}$ at the remaining modes $k' > k$ have to be shifted accordingly. (This process is referred to as quantum jumps in some of the literature on quantum diffusion \cite{Wiseman1993, Brun2001}; the only difference here is that the observer acts as the bath.) This gradual updating of $\vectau_{o}$ based on the history of outcomes $\vecmu$ means that, in effect, measurement optimization can operate not only in Hilbert \textit{space}, but also in \textit{time}. As the gradual collapse takes place, $\vectau_{o}$ and $\vecmu$ will grow in tandem, with $\vecmu$ lagging behind $\vectau_{o}$ by one element. In the next section we shall model the relationship between the optimal parameters $\vectau_{o}$ and the history of outcomes $\vecmu$ with a simple Bayesian network that can be used to infer the identity of the candidate states.

%%%%%%%%%%%%%%%%%%%%%%%%%%%%%%%%%%%%%%%%%%%%%%%%%%%%%%%%%%%%%%%%%%%%%
\section{Adaptive measurement as an algorithmic problem}
\label{sec:Algorithmic}

Although Eqs. (\ref{eq:SimilarityGeneralized}) and (\ref{eq:ArgmaxGeneralized}) are self-contained, solving them for $\vectau_{o}$ analytically is non-trivial. It may therefore be more practical to resort to numerical methods. These methods simulate all possible combinations of outcomes $\vecmu$ and, for each such combination, perform a parameter sweep over the elements of $\vectau$ so as to maximize the figure of merit $\mathcal{F}$ of interest. We shall see how this can be tackled by different algorithms which can be either local-optimal or global-optimal, depending on the development cost one is willing to allocate to the problem. These algorithms have in common that they give shape to Bayesian networks which can subsequently be used by experimentalists as look up  tables whereby each setting $\tau^{(k+1)}$ is linked to the history of outcomes $\vecmu^{(k)} = \hak{\mu^{(1)},\cdots,\mu^{(k)}}$. Let us first introduce the notion of superoperator with which the networks are most conveniently traversed.

\subsection{Generalized measurements as superoperators}

\begin{figure}[h]
  \centering\includegraphics[width=.7\columnwidth]{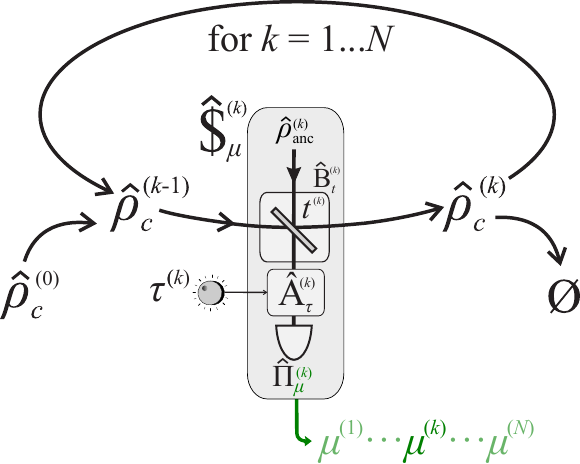}
  \caption{Representation of the generalized measurement setup of Fig. \ref{fig:Setup} as a recursive two-mode superoperation. The superoperator could be thought of as a black box that spans the Hilbert space $\mathcal{H}^{(0)}$ of the input state. This black box features a classical control knob for $\tau$, a classical readout for $\mu$, as well as a quantum input port for $\rhohat_{c}^{(k-1)}$ which is output as $\rhohat_{c}^{(k)}$. The transmission $t^{(k)}$ of the beam splitter $\Bhat_{t}^{(k)}$ can in principle be incorporated into $\tau^{(k)}$ as yet another unitary degree of freedom.}
	\label{fig:RecursionRepresentation}
\end{figure}

We have concluded Sec. \ref{sec:Algebraic} by noting that, under a gradual collapse of the wave function, one can further exploit the adaptation of the parameters $\vectau_{o}$ to the history of outcomes $\vecmu$. Gradual collapse is however unwieldy to treat with multi-modal matrices such as those of Eq. ($\ref{eq:GeneralizedBorn}$) as that would require the cumbersome nesting of partial traces. A better solution, which readily lends itself to implementation, is to confine the whole problem to the Hilbert space of the zeroth mode and recursively update the candidate states and their probability distributions upon each collapse of the ancillary mode. This recursion, schematized in Fig. \ref{fig:RecursionRepresentation}, transforms the candidate states from one collapse to the next via a superoperator $\sop_{\mu}$ \cite{Preskill1998}:
\beqa
\rhohat_{c}^{(k)} & = & \sop_{\mu}^{(k)}\hak{ \rhohat_{c}^{(k-1)} }\nonumber\\
& = & \sum\limits_{i,n,m=0}^{\infty} \bra{n}\rhohat_{\mbox{\scriptsize aux}}^{(k)}\ket{m} \, \Khat_{i,n} \, \rhohat_{c}^{(k-1)} \, \Khat^{\dag}_{m,i}, 
\label{eq:Superoperator}
\eeqa
where the Kraus operators are given by
\beq
\Khat_{i,j}  = \bra{i} \tes{\hat{\mathbb{I}} \otimes \sqrt{\Pihat_{\mu^{(k)}}} \Ahat_{\tau^{(k)}}} \Bhat_{t^{(k)}} \ket{j}.
\label{eq:Kraus}
\eeq
Note that in both Eqs. (\ref{eq:Superoperator}) and (\ref{eq:Kraus}), the Dirac notation pertains to the ancillary mode. A detailed derivation of the superoperation is given in Sec. \ref{sec:AppendixSuperoperators}. 

The multi-mode expression for Born's rule, Eq. ($\ref{eq:GeneralizedBorn}$), can be re-expressed as $N$ recursive superoperations whereby the probability distributions up to the $k$th ancillary mode collapse are given by
\beq
p_c^{(k)}(\vecmu^{(k)}) = p_c^{(0)} \cdot p_{\vecmu^{(k)}\mid c}^{(k)},
\eeq
where
\beqa
p_{\vecmu^{(k)}\mid c}^{(k)} & = & \mbox{Tr}\braces{\sop_{\mu}^{(k)}\hak{\sop_{\mu}^{(k-1)}\hak{\cdots\sop_{\mu}^{(1)}\hak{\rhohat_{c}^{(0)}}}}}.
\label{eq:RecursiveSuperopProb}
\eeqa

\subsection{The tree data structure}
\label{sec:TreeDataStructure}

\begin{figure}[h]
  \centering\includegraphics[width=1\columnwidth]{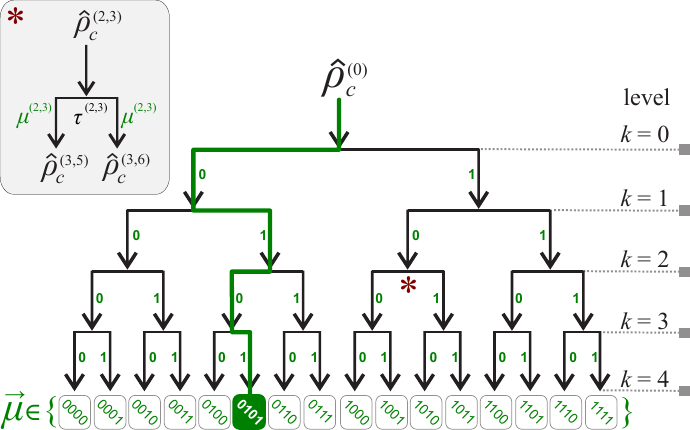}
  \caption{In a gradual collapse scenario, the relationship between the history of outcomes and the unitary parameters is best represented by a tree structure where each ramification corresponds to the possible outcomes of the direct measurement. At each node, one needs to determine the parameters $\tau_{o}^{(k)}$. (The entire data structure is in this sense a \textit{decision} tree.) Here, we show the case of a direct measurement which only has $M = 2$ possible outcomes---e.g., an avalanche photo diode or a homodyne detector whose quadrature readouts are partitioned into two complementary ranges. The candidate states undergo $N = 4$ de-localizations such that, in the end, the whole setup presents $M^N = 16$ outcomes. Any given run will traverse one these 16 distinct branches $\vecmu_l$ with a probability of $\sum_{c} p_{c}(\vecmu_{l})$. In this case we have highlighted the outcome $l = 6$ whose sequence is $\vecmu_{6} = [0,1,0,1]$. With such an \textit{a posteriori} knowledge, one can then infer backwards what state $c$ was most likely to have been input. In order to illustrate the labeling in our notation, the inset presents the node marked by an asterisk.}
	\label{fig:TreeRepresentation}
\end{figure}

We now have all the tools to present the Bayesian network that governs the probability distributions $p_{c}^{(k)}$ under a gradual collapse scenario. This network is best represented by a class probability tree \cite{Murthy1998}. The tree data structure is indeed an increasingly  favored choice in the represention of the  information-theoretic flow in many quantum processes (cf. Refs. \cite{Hentschel2010, Childs2013, Mancinska2013, Pozza2014, Oi2014, Wang2014}). In our case, each node in the tree stores the density matrices of the candidates states $\rhohat_{c}^{(k-1)}$, their probabilities $p_{c}^{(k-1)}$, and the values of the unitary parameters $\tau^{(k)}$ and $t^{(k)}$. From each node emanate $M$ edges which correspond to the $M$ possible outcomes of the direct measurement. This pattern is iterated all the way to the depth $N$ of the tree, where each leaf node takes on a unique label $l \in \braces{1, \cdots, M^N}$. The root node, which contains the input candidate states and their prior probabilities, lies at level $k = 0$. Finally, the sequence of outcomes $\vecmu_{l}$ leading up to the $l$th leaf is called a branch. The tree data structure is illustrated for the case of $N = 4$ and $M = 2$ in Fig. \ref{fig:TreeRepresentation}.

Note that in order not to overload the notation, we have not specified any label to uniquely identify the nodes within a given level. Notational rigor should however require all parameters pertaining to any given node to be labeled by the coordinates $(k,\nu)$ where $k\in\braces{1,\cdots,N}$ is the level in the tree and $\nu\in\braces{1,\cdots,M^k}$ is the horizontal position of the node at that level. This more rigorous notation is exemplified in the inset of Fig. \ref{fig:TreeRepresentation}.

Ideally, the goal of a generalized measurement is to have any given leaf mapped by exactly one candidate state $c$, or at least to minimize the overlap of the projected probability distributions at the leaves. Only then will the figures of merit $\mathcal{F}$ be maximized. (We shall return to the exact definitions of the figures of merit in the next section.) As we saw in the previous section, we can determine the sequence $\vectau$ which satisfies Eqs. (\ref{eq:SimilarityGeneralized}) or (\ref{eq:ArgmaxGeneralized}). In this case, all the unitary parameters in the nodes at a given level $k$ will be equal. Although this represents an improvement over the direct measurement, it does not take advantage the chronology of the outcomes $\mu^{(1)},\cdots,\mu^{(k-1)}$. For that, even within a given level, the parameters $\tau^{(k)}$ at each node will have to be adapted to the particular shape (in Hilbert space) that the candidate states have inherited from previous measurement. Let us next explain how the parameters $\tau^{(k)}$ can be optimized for each node.

\subsection{Optimization algorithms}
\label{sec:OptimizationAlgorithms}

One of the simplest methods determines the optimal parameters $\tau_{o}^{(k)}$ on the fly: At each $k$th partial measurement, it tries to maximize the figure of merit by performing a parameter sweep over $\tau^{(k)} \in \mathcal{T}$. (This is known as a greedy algorithm \cite{CLRS2009} whereby optimization abides by a short-term, maximum-gain policy.) In practice, one could proceed with a so-called pre-order traversal: One first determines the parameter $\tau_{o}^{(k)}$ at the current node and then recursively visits the children nodes from, say, left to right. At the end, all internal nodes will have been assigned a value for $\tau_{o}^{(k)}$. In an experimental context,  each time a partial measurement $\mu^{(k)}$ is recorded, $\tau_{o}^{(k+1)}$ is looked up from the tree and fed-forward to the next measurement. As the sequence of outcomes $\vecmu$ is gradually acquired, one could even perform a ``live update'' on the confidence of having identified a certain state $c$ with a maximum likelihood estimation based on the probabilities $p_{\vecmu^{(k)} \mid c}$.

Though relatively simple to program and often as efficient as global optimization methods \cite{Murthy1998}, greedy algorithms run the risk of getting stuck at local optima. Indeed, the parameters  $\tau_{o}^{(k)}$ are determined locally at each node in a top-down manner from root to leaf. A truly global algorithm, on the other hand, would not just perform a parameter sweep over the parameter range $\mathcal{T}$ one node at a time, but would probe all possible combinations of $\tau$ over all $\frac{M^N-1}{M-1}$ internal nodes. Due to the exponential growth of the tree with its depth $N$, such a global parameter sweep is likely to be numerically demanding. Among the global optimization methods, dynamic programming is one of the most tractable candidates \cite{CLRS2009, Bouten2009, Pozza2014} as it avoid redundancies in the parameter sweep of the decision trees while still probing all combinations. Several hybrid heuristics also exist which combine global- and local-optimal performance. Many of the established techniques from machine learning  could be relevant in this regard \cite{Bennett1994, Murthy1998}. One such technique---particle swarm optimization---has incidentally been applied to phase estimation by Hentschel and Sanders in Ref. \cite{Hentschel2010} 

%%%%%%%%%%%%%%%%%%%%%%%%%%%%%%%%%%%%%%%%%%%%%%%%%%%%%%%%%%%%%%%%%%%%%
\section{Figures of merit and simulations}
\label{sec:Simulations}

In order not to clutter the previous discussions, we have so far only referred to the figures of merit symbolically as $\mathcal{F}$. Let us now define some of them in detail so as to examine the performance trends of generalized measurement. 

\subsection{Figures of merit}
\label{sec:FiguresOfMerit}

As we briefly mentioned in the introduction, a measurement can always be reduced to a discrimination problem: If we measure some state $\rhohat \in \mathcal{H}$ and completely characterize it, we are in a sense discriminating it from everything else in $\mathcal{H}$ that it is not. The pool of candidate states is in that case infinite. If we however discretize the pool of candidate states, we are back to the discrimination problem we have treated so far. We shall therefore surmise that an optimal measurement is that which best distinguishes the elements in a given pool of possibilities. Experimentally, the only evidence we have to go by when distinguishing quantum states $c$ is the probability distribution functions $p_c(\vecmu)$ they cast on the measurement spectrum. It is then a natural choice to start with figures of merit $\mathcal{F}$ that depend on $p_c(\vecmu)$.

\subsubsection{Distinguishability}

Fuchs \cite{Fuchs1996, Fuchs1999} provides a thorough survey of discrimination measures. Let us adapt some of them to our purposes. The first, which we shall generically refer to as ``distinguishability'' and denote by $\mathcal{D}$, is based on the Bhattacharyya coefficient \cite{Bhattacharyya1943, Bhattacharyya1946, Laghaout2015}
\beq
\mbox{BC}(p_{c}, p_{c'}) = \sum_{x} \sqrt{p_{c}(x) \, p_{c'}(x)} \in \hak{0,1},
\label{eq:BC}
\eeq 
between two normalized probability distributions $p_{c}(x)$ and $p_{c'}(x)$. This coefficient, which is just an inner product of two functions of $x$, quantifies their similarity in the same way that a dot product quantifies the overlap of two vectors. In our case, we are dealing with a pool of $C$ different probability distributions. We therefore propose to define a Bhattacharyya coefficient which averages out the similarity between all the possible pairs in the pool
\beq
\mbox{BC}(p_{1},\cdots,p_{C}) = \sum\limits_{\substack{c,c' = 1\\c \neq c'}}^C p_{c}^{(0)} p_{c'}^{(0)} \mbox{BC}(p_{c}, p_{c'}).
\label{eq:AveragedBC}
\eeq 
Note that the averaging took into account the prior probabilities $p_{c}^{(0)}$ of the candidate states. Since we are interested in distinguishability rather than similarity, we shall use a modification of the Bhattacharyya coefficient referred to as the Hellinger distance and defined as
\beq
\mbox{HD}(p_{i}, p_{j}) = \sqrt{1-\mbox{BC}(p_{i}, p_{j})}.
\label{eq:HD}
\eeq
Putting together Eqs. (\ref{eq:AveragedBC}) and (\ref{eq:HD}), we define distinguishability for our purposes as
\beq
\mathcal{D} = \sqrt{1- \sum\limits_{\substack{c,c' = 1\\c \neq c'}}^C p_{c}^{(0)} p_{c'}^{(0)} \sum\limits_{l=1}^{M^N} \sqrt{p_{c}(\vecmu_{l}) \cdot p_{c'}(\vecmu_{l})}}.
\label{eq:Distinguishability}
\eeq
It should be clear that although we have designed this figure of merit to be as comprehensive as possible, it is by no means better than any of the other ones described in Refs. \cite{Fuchs1996, Fuchs1999}. 

\subsubsection{Mean min-to-max ratio}

Another figure of merit which quantifies the overlap of two probability distributions is the mean ratio of the minimum-to-maximum probability distributions
\beq
\mathcal{R} = \sum\limits_{l=1}^{M^N} \sum\limits_{c = 1}^{C} p_{c}(\vecmu_{l}) \frac{\min\limits_{c'} \braces{p_{c'}(\vecmu_{l})}}{\max\limits_{c''} \braces{p_{c''}(\vecmu_{l})}}.
\eeq
Unlike $\mathcal{D}$, which is to be maximized, we should aim to minimize $\mathcal{R}$. This definition of the mean ratio of minimum-to-maximum probabilities, though intuitive, cannot be easily applied to cases where $C \ge 3$. We shall therefore only apply to pools of candidate states containing two elements only.

\subsubsection{Error probability}

Both $\mathcal{D}$ and $\mathcal{R}$ stem from an algebraic, rather than an operational, rationale. A more operational figure of merit would be the discrimination error $\mathcal{E}$, i.e., the probability of mistaking one state $c$ for another $c' \neq c$ and vice versa. Just like $\mathcal{R}$, $\mathcal{E}$ should be minimized. It is given by 
\beq
\mathcal{E} = p_{c}^{(0)} P_{c'}(c) + p_{c'}^{(0)} P_{c} (c'), 
\label{eq:DiscErr}
\eeq
where $P_{c}(c')$ is the probability that the generalized measurement identifies state $c'$ as $c$ \cite{Ferraro2005}. Ideally, we should have $P_{c}(c) = P_{c'}(c') = 1$ and $P_{c}(c') = P_{c'}(c) = 0$ for $c \neq c'$. These identification probabilities are given by
\beq
P_{c}(c') = \sum\limits_{l \in \mathcal{L}_{c}} p_{\vecmu_l \mid c'}^{(N)}
\eeq
where $p_{\vecmu_l \mid c'}^{(N)}$ is expressed in Eq. (\ref{eq:RecursiveSuperopProb}). $\mathcal{L}_{c}$ indicates the set of leaves $l$ most likely to be attained by state $c$. In other words, if state $c$ is most likely to traverse branch $\vecmu_{l}$, then we assign leaf $l$ to $\mathcal{L}_{c}$. If we limit ourselves to two candidate states $\rhohat_1$ and $\rhohat_2$, we will have
\beqa
l \in \mathcal{L}_{1} & \Leftrightarrow & p_{1}(\vecmu_l) > p_{2}(\vecmu_l), \label{eq:L1set}\\
l \in \mathcal{L}_{2} & \Leftrightarrow & p_{2}(\vecmu_l) > p_{1}(\vecmu_l). \label{eq:L2set}
\eeqa
If $p_{1}(\vecmu_l) = p_{2}(\vecmu_l)$, the sequence $\vecmu_{l}$ will correspond to an inconclusive outcome. Note that, for completeness, $\mathcal{L}_{1}$ and $\mathcal{L}_{2}$ are non-overlapping and their cardinalities add up to $M^N$, i.e., the total number of leaves. Furthermore, although we have only expressed $\mathcal{E}$ for the case of two candidate states only, it can easily be generalized to an average of pairwise discrimination errors, in a similar fashion to what we did with the Bhattacharyya coefficient in Eq. (\ref{eq:AveragedBC}).

\subsubsection{Bell factor}

Quantum measurements are not merely ends in themselves, but are often part of a broader process or computation. In fundamental physics, investigations of nonlocality are one such application. Although Bell tests are based on measurement, it is not the identification or the discrimination of quantum states that is their primary purpose. Rather, it is the Bell factor $\mathcal{B}$, a statistical discrepancy between the classical and quantum correlations of space-like events, which is the prime figure of merit. Though indirectly related, $\mathcal{B}$ and $\mathcal{E}$ are not necessarily optimized by the same configuration of measurement. The Bell factor we shall consider is that which can be produced by a tri-partite W state, which we shall express in Fock basis as
\beq
\ket{\mbox{W}} = \frac{1}{\sqrt{3}}\tes{\ket{0,0,1} + \ket{0,1,0} + \ket{1,0,0}}. \label{eq:Wstate}
\eeq
The full details of the Bell setup and the exact expression for $\mathcal{B}$ are provided in Ref. \cite{Laghaout2011} so we shall not reproduce them here. It suffices to know that the Bell test rests on our ability to distinguish, at one or more modes of the W state, the two candidate states 
\beq
\rhohat_{1} \equiv \frac{\ket{0}+\ket{1}}{\sqrt{2}} \mbox{~and~}  \rhohat_{2} \equiv \frac{\ket{0}-\ket{1}}{\sqrt{2}},
\label{eq:OurQubits}
\eeq
whose probabilities of incidence are equal $p_{1}^{(0)} = p_{2}^{(0)} = \frac{1}{2}$. 

Ideally, this discrimination is done by the POVMs
\beq
\Ohat_1 = \begin{bmatrix}
	\frac{1}{2} & \frac{1}{2} \\[0.3em]
	\frac{1}{2} & \frac{1}{2}
	\end{bmatrix}, \mbox{~and~} \Ohat_2 =  \begin{bmatrix}
	\frac{1}{2} & -\frac{1}{2} \\[0.3em]
	-\frac{1}{2} & \frac{1}{2}
	\end{bmatrix}.
\eeq
In practice, however, these ideal projectors can at best be approximated with the superoperators we obtained from our generalized measurement. This is achieved by bundling all the sequences of superoperations that are most likely to be projected on by a given state such that
\beqa
\Ohat_1 \approx \sum\limits_{l \in \mathcal{L}_1} \sop_{\vecmu_{l}} \hak{\rhohat_{1}}, \mbox{~and~} \Ohat_2 \approx \sum\limits_{l \in \mathcal{L}_2} \sop_{\vecmu_{l}} \hak{\rhohat_{2}},
\eeqa
where $\mathcal{L}_1$ and $\mathcal{L}_2$ were defined in Eqs. (\ref{eq:L1set}) and (\ref{eq:L2set}). $\sop_{\vecmu_{l}}$ represents the $N$ nested superoperations along branch $\vecmu_l$:
\beq
\sop_{\vecmu_{l}} = \sop_{\mu}^{(N)} \cdots \sop_{\mu}^{(k)} \cdots \sop_{\mu}^{(1)}.
\eeq

Finally, let us recall that for the particular case of the W state, a positive $\mathcal{B}$ is indicative of a violation of Bell's inequality and the larger its positive amplitude, the more decisive is the violation.

\subsubsection{Orthogonality}

Let us conclude the presentation of figures of merit by introducing the orthogonality $\Omega$ of two states $\rhohat_1$ and $\rhohat_2$, which we define as the complement of fidelity
\beq
\Omega(\rhohat_1, \rhohat_2) = 1-\mbox{Tr}\braces{\rhohat_1 \rhohat_2}.
\eeq

Some previous work, particularly by Takeoka \textit{et al.}, has presented orthogonality as a central criterion for the adaptive discrimination of quantum states \cite{Walgate2000, Acin2005, Takeoka2005, Takeoka2006}. If we assume that all the candidate states in the pool are initially orthogonal $\Omega(\rhohat_{i}^{(0)}, \rhohat_{j}^{(0)}) = 1, \, \forall i \neq j$, it indeed makes sense to insure they remain orthogonal after $k$ partial measurements such that $\Omega(\rhohat_{i}^{(k)}, \rhohat_{j}^{(k)}) = 1, \, \forall k > 0$. Any non-orthogonality that is acquired during the gradual collapse of the states will represent a fundamental and irrecoverable loss of distinguishability. Though intuitive, the requirement of orthogonality does not have any operational motive in itself. In fact, it leads to singularities in the optimization of the unitary parameters $\tau^{(k)}$ if the candidate states are the Fock qubits of Eq. (\ref{eq:OurQubits}) \cite{Takeoka2006}. The utility of orthogonality as a figure of merit shall therefore be de-emphasized in our simulations.

\subsection{Numerical simulations}
\label{sec:NumSims}

In this section, we shall present some trends in the figures of merit $\mathcal{D}$, $\mathcal{B}$, $\mathcal{R}$ and $\mathcal{E}$ under various generalized measurement configurations. The pool of candidate states consists of $C$ qubits equally spread along the real-valued longitudinal cross-section of the Bloch sphere such that
\beq
\rhohat_{c}^{(0)} = \begin{bmatrix}
	\cos\tes{\frac{\theta}{2}}^2 & \cos\tes{\frac{\theta}{2}}\sin\tes{\frac{\theta}{2}} \\[0.3em]
	\sin\tes{\frac{\theta}{2}}\cos\tes{\frac{\theta}{2}} & \sin\tes{\frac{\theta}{2}}^2
	\end{bmatrix}, \mbox{~and~} p_{c}^{(0)} = \frac{1}{C}, \label{eq:QubitPool}
\eeq
where $\theta = \frac{(2c-1)\pi}{C}$ and $c\in\braces{1,\cdots,C}$. In the special case of $C = 2$, we are left with the two states of Eq. (\ref{eq:OurQubits}).

Two types of operations $\Ahat_{\tau}$ shall be considered. The first is a Hadamard rotation, defined in Fock space as 
\beq
\AHad_{\tau} = \begin{bmatrix}
	\cos\tes{\frac{\tau}{2}}^2 & \cos\tes{\frac{\tau}{2}}\sin\tes{\frac{\tau}{2}} \\[0.3em]
	\sin\tes{\frac{\tau}{2}}\cos\tes{\frac{\tau}{2}} & \sin\tes{\frac{\tau}{2}}^2
	\end{bmatrix}, \label{eq:HadamardGate}
\eeq
where $\tau \in \mathcal{T} = \hak{-\pi, \pi}$. The second operation is a coherent displacement 
\beq
\ADis_{\tau} = e^{\tau \adag - \tau^{*}\ahat}, \label{eq:CoherentDisplacer}
\eeq
where $\ahat$ and $\adag$ are the creation and annihilation operators, respectively. (An explicit expression for $\Ahat_{\tau}^{\mbox{\scriptsize (Dis.)}}$ acting on Fock states is given in Refs. \cite{Kral1990, Moya-Cessa1995}.) The parameter range for the displacements shall be confined to the real segment $\mathcal{T} = \hak{-1, 1}$.

Because the simulations are numerical, we only probe the parameter ranges $\mathcal{T}$ at a finite number $S_{\mathcal{T}} \in \mathbb{N}$ of sample points. We chose $S_{\mathcal{T}}^{\mbox{\scriptsize Had.}} = 40$ and $S_{\mathcal{T}}^{\mbox{\scriptsize Dis.}} = 10$ equidistant points for the Hadamard angle and coherent displacement, respectively. Furthermore, in the search for the optimal parameter $\tau_{o}$ at any given mode, the density of the initial mesh of parameters is doubled (i.e., $S_{\mathcal{T}} \rightarrow 2\times S_{\mathcal{T}}$) so long as the figure of merit does not converge at a certain satisfactory rate.

Three types of von Neumann projections $\Pihat_{\mu}$ shall be simulated. The first is an imperfect on-off click detector such as an avalanche photo diode (APD). The two possible outcomes ($M = 2$), no-click and click, are given in Fock space by 
\beqa
\PiAPD_{1} & = & \sum\limits_{n=0}^{\infty} \tes{1-\eta}^{n} \ketbra{n}{n}, \mbox{~and} \\
\PiAPD_{2} & = & \id - \PiAPD_{1},
\eeqa
respectively, where $\eta \in \hak{0, 1}$ is the quantum efficiency \cite{Ferraro2005}. A fine-grained generalization of the APD is the photon number resolving detector (PNRD),
\beqa
\PiPNRD_{\mu \le \nsat} & = & \eta^{\mu-1} \sum\limits_{n=\mu-1}^{\infty} \tes{{1-\eta}}^{{n-\mu + 1}} {k \choose {\mu - 1}} \ketbra{n}{n}, \nonumber\\\\
\PiPNRD_{\mu = \nsat + 1} & = & \id - \sum\limits_{\mu' = 1}^{\nsat} \PiPNRD_{\mu'},
\eeqa
where $\nsat$ is the photon count at which the PNRD saturates such that $M = \nsat + 1$. (An APD is a PNRD which already saturates at $\nsat = 1$ in that it cannot differentiate one photon from more than one.) The last von Neumann projector we shall use is a homodyne detector (HD) which bins the real quadratures $x$ into positive and negative values
\beqa
\PiHD_{1} & = & \int_{-\infty}^{0} \! \ketbra{x}{x} \, \mathrm{d}x, \label{eq:PiHD1}\\
\PiHD_{2} & = & \int_{0}^{\infty} \! \ketbra{x}{x} \, \mathrm{d}x. \label{eq:PiHD2}
\eeqa

Finally, we shall set the multi-modal beam splitter $\Bhattilde$ to de-localize the candidate states equally into $N$ modes such that the transmission of the zeroth mode at the $k$th splitting is given by
\beq
t^{(k)} = \sqrt{\frac{N - k}{N - k + 1}}. \label{eq:Transmission}
\eeq
For simplicity, all ancillary modes shall remain empty by setting $\rhohat_{\mbox{\scriptsize anc}}^{(k)} = \ketbra{0}{0}$, $\forall k$.

Note that unless stated otherwise, all the results presented below were obtained using greedy algorithms whereby the distinguishability $\mathcal{D}$ was the figure of merit optimized at each node.  (Other figures of merit could of course have been chosen to drive the optimization, depending on the application at hand.) This arbitrary choice, in addition to the finite sampling of the parameter ranges $\mathcal{T}$, means that the figures of merit presented here are under-estimated and that the implementation of a global optimization algorithm with a denser parameter sampling is bound to demonstrate better performances.

\subsubsection{Hadamard rotation and APD}

\begin{figure*}
\begin{tabular}{ c c c c}
\multicolumn{4}{c}{\includegraphics[width=.45\textwidth]{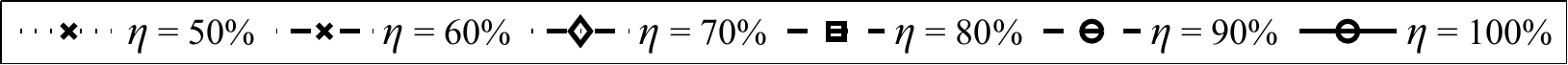}} \\
\includegraphics[width=0.250\textwidth]{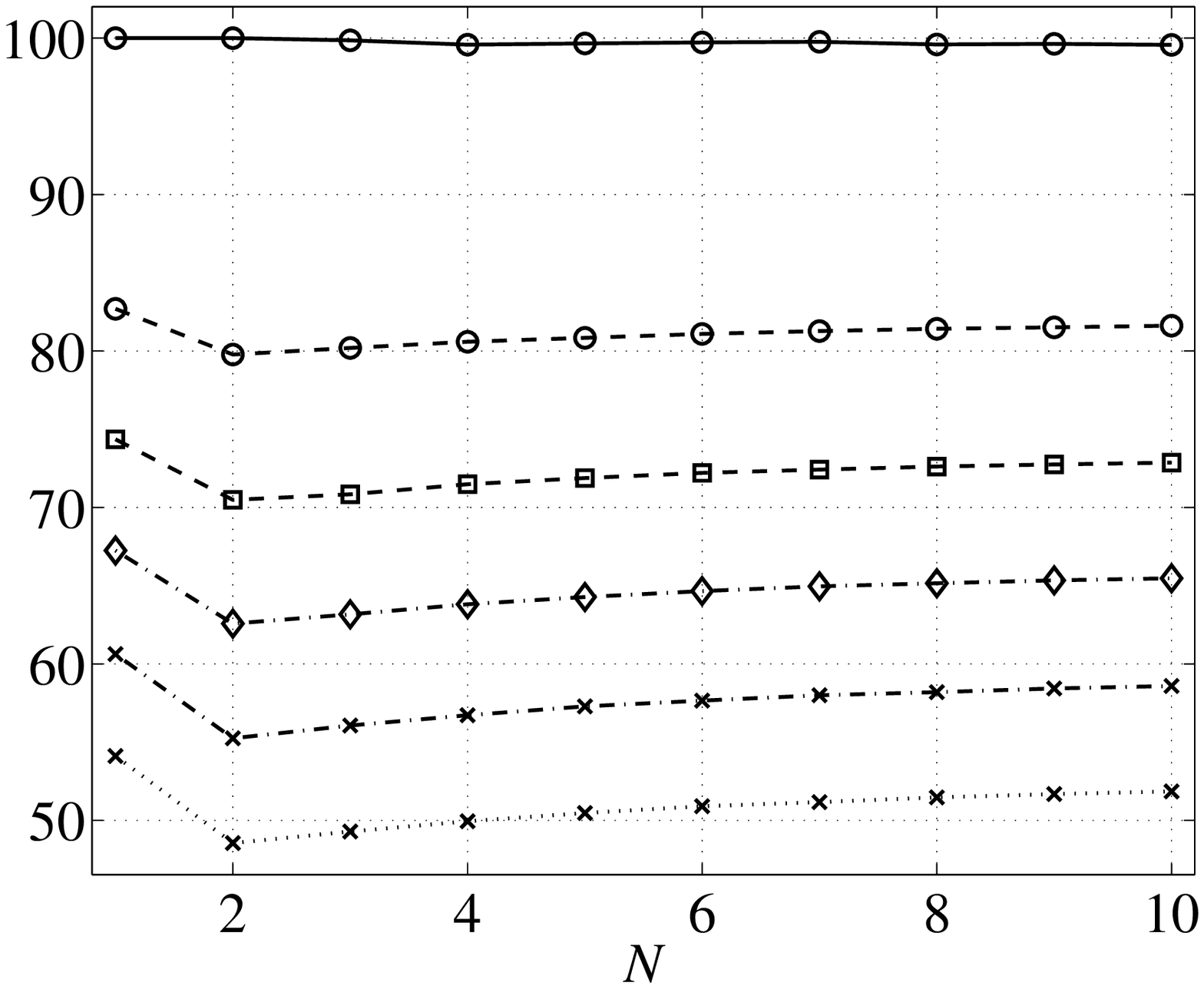} & 				\includegraphics[width=0.250\textwidth]{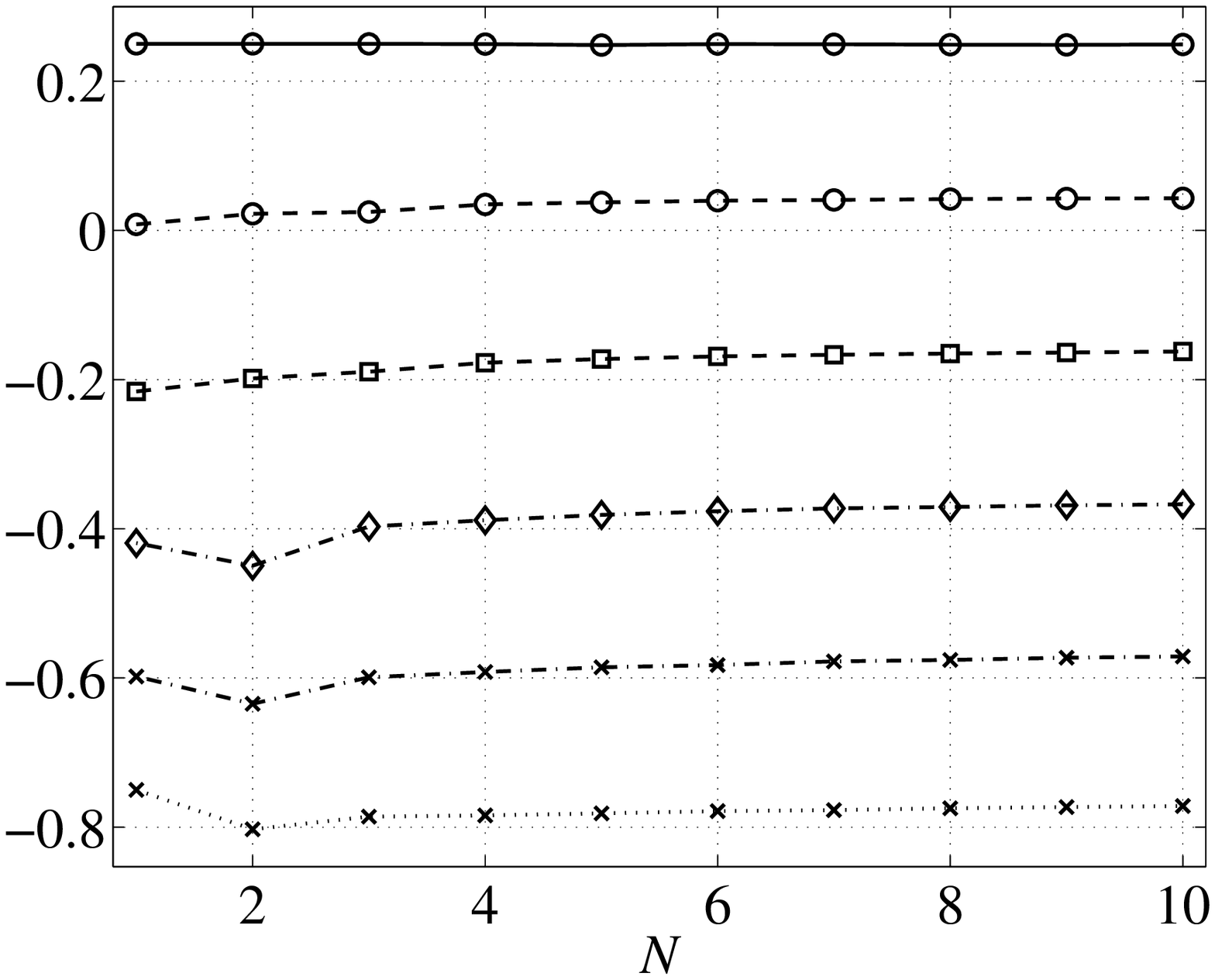} & \includegraphics[width=0.250\textwidth]{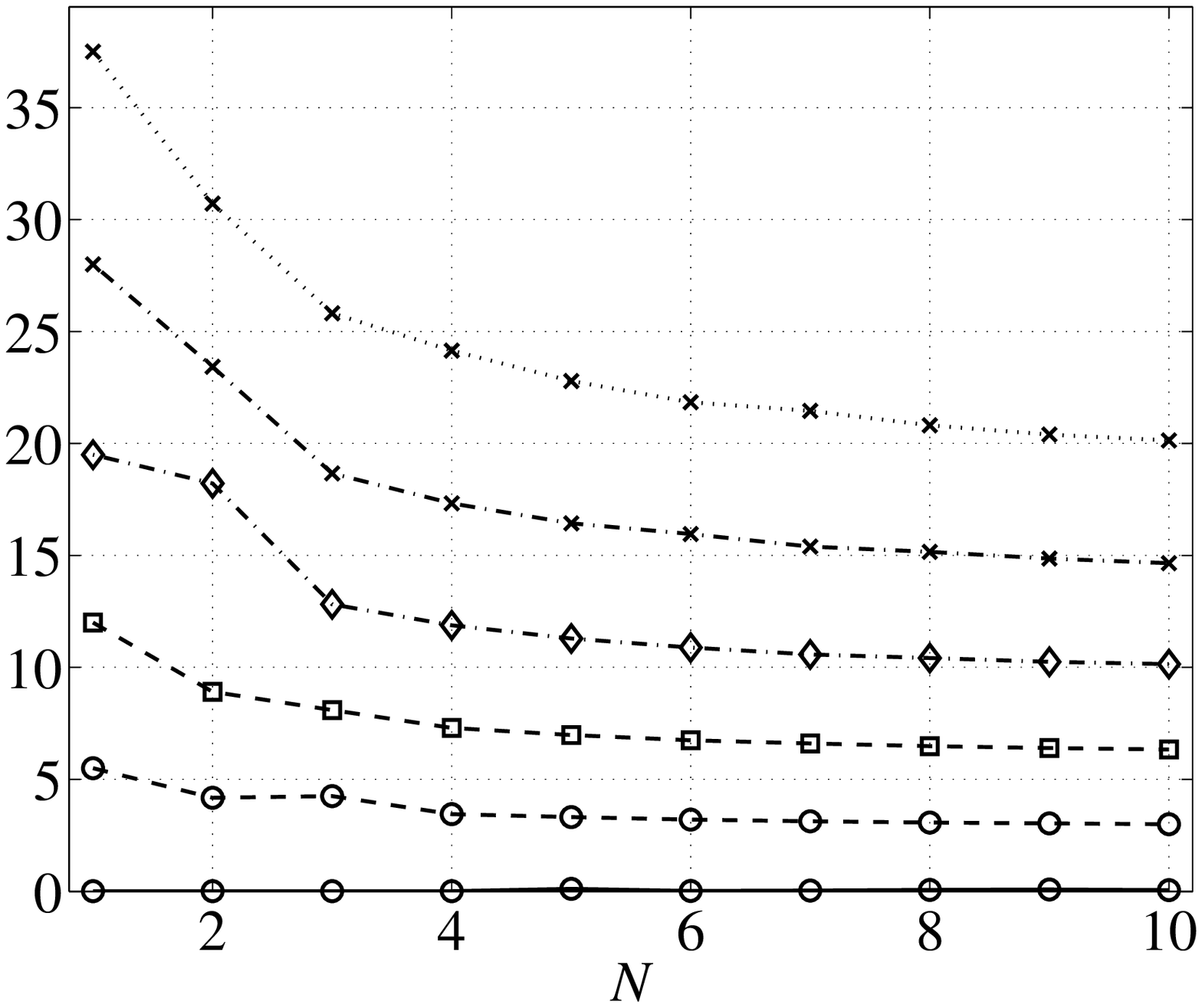} & \includegraphics[width=0.250\textwidth]{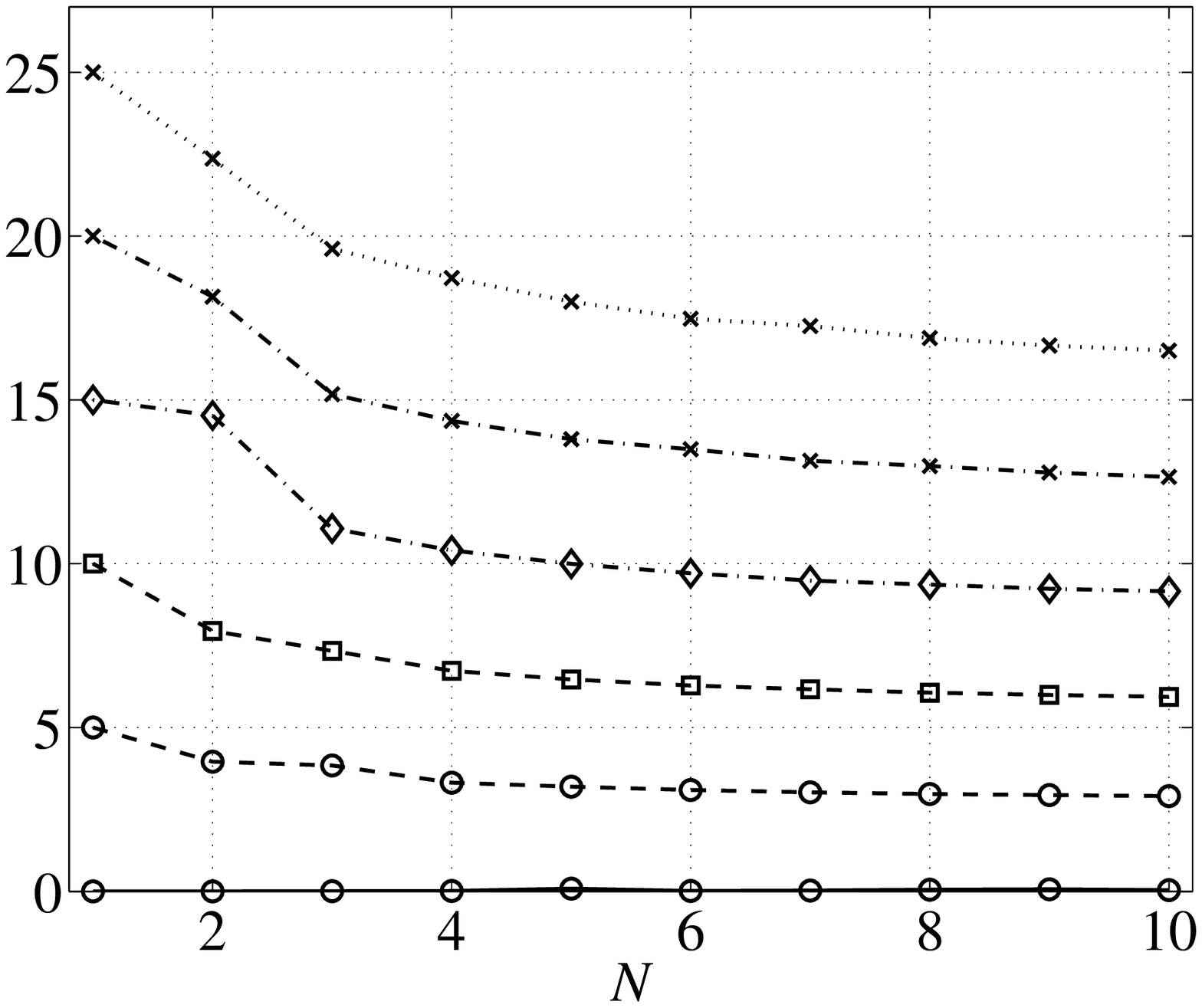} \\ 
\scriptsize {\scriptsize Distinguishability $\mathcal{D}$ [\%]} & {\scriptsize Bell factor $\mathcal{B}$}& {\scriptsize Mean min-to-max ratio $\mathcal{R}$ [\%]} & {\scriptsize Discrimination error $\mathcal{E}$ [\%]}   
\end{tabular}
\caption{The figures of merit as a function of $N$ using a Hadamard rotator $\AHad$ and an APD $\PiAPD$ with different quantum efficiencies $\eta$. The pool of candidate states consists of the two qubits of Eq. (\ref{eq:OurQubits}).}
\label{fig:HG_APD}
\end{figure*}

Let us for now set $C = 2$ so that the candidate states are those of Eq. (\ref{eq:OurQubits}). The first combination we shall simulate comprises $\AHad_{\tau}$ and $\PiAPD_{\mu}$ at each partial measurement (Fig. \ref{fig:HG_APD}). Under a direct measurement setting $N = 1$, the particularity of this combination is that it satisfies \textit{exactly} the similarity transformation (\ref{eq:SimilarityDirect}) provided we set $\tau_{o} = \pm\frac{\pi}{2}$ and $\eta = 100\%$:
\beq
\sqrt{\AHad_{\tau_{o}}{}^{\dagger} \, \PiAPD_{\mu} \, \AHad_{\tau_{o}}} = \Ohat_{\mu}, \, \forall \mu.
\eeq
Such a combined effect of $\AHad_{\pm{\pi/2}}$ and $\PiAPD_{\mu}$ is thus equivalent to that of an ideal POVM and there would therefore be no need to resort to a generalized measurement with $N \ge 2$. This can be seen in Fig. \ref{fig:HG_APD} where all the figures of merit with $\eta = 100\%$ are readily saturated at their global optima and do not exhibit any $N$-dependence. In practice, however, not only does there not exist any trivial laboratory setup which implements $\AHad_{\pm{\pi/2}}$, but even if there were, the quantum efficiency of the APD will likely be much less than unity. This non-unit quantum efficiency causes a degradation in all figures of merit. For example, simply going from $\eta = 100\%$ to $\eta = 90\%$ decreases the Bell factor from its maximum at $\mathcal{B} = 0.25$ to an inconclusive $\mathcal{B} = 0$ while the discrimination error $\mathcal{E}$ jumps from zero to $5\%$. With this in mind, it is interesting to see whether a generalized measurement can compensate for the lower quantum efficiency. This turns out to indeed be the case: The overall trends of all figures of merit improves with an increased $N$. For instance, the discrimination error of a generalized measurement with $N = 5$ and $\eta = 70\%$ is as good as a direct measurement $N = 1$ with a higher $\eta = 80\%$. 

\subsubsection{Coherent displacement and APD}

\begin{figure*}
\begin{tabular}{ c c c c}
\multicolumn{4}{c}{\includegraphics[width=.45\textwidth]{legend.pdf}} \\
\includegraphics[width=0.250\textwidth]{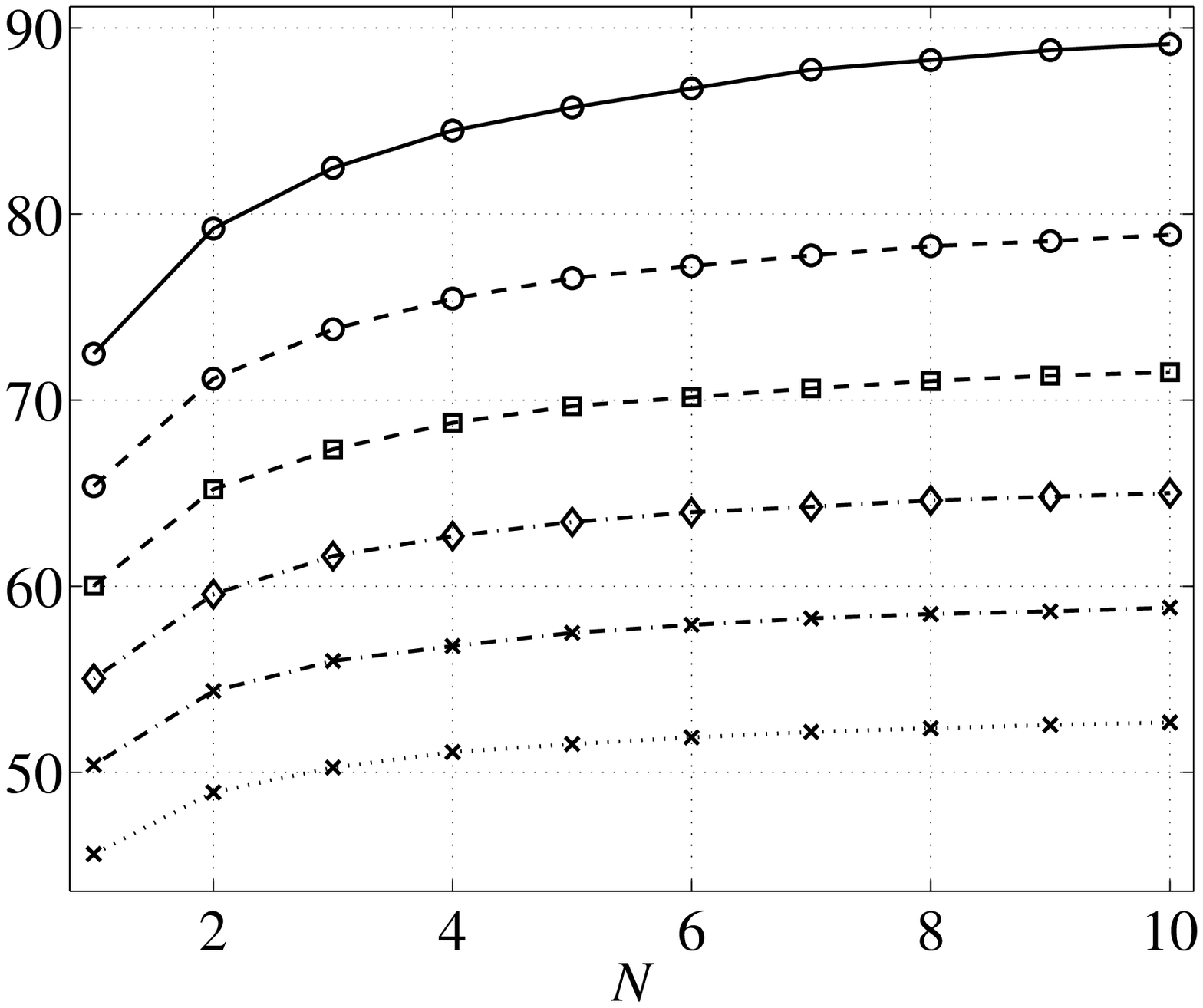} & 				\includegraphics[width=0.250\textwidth]{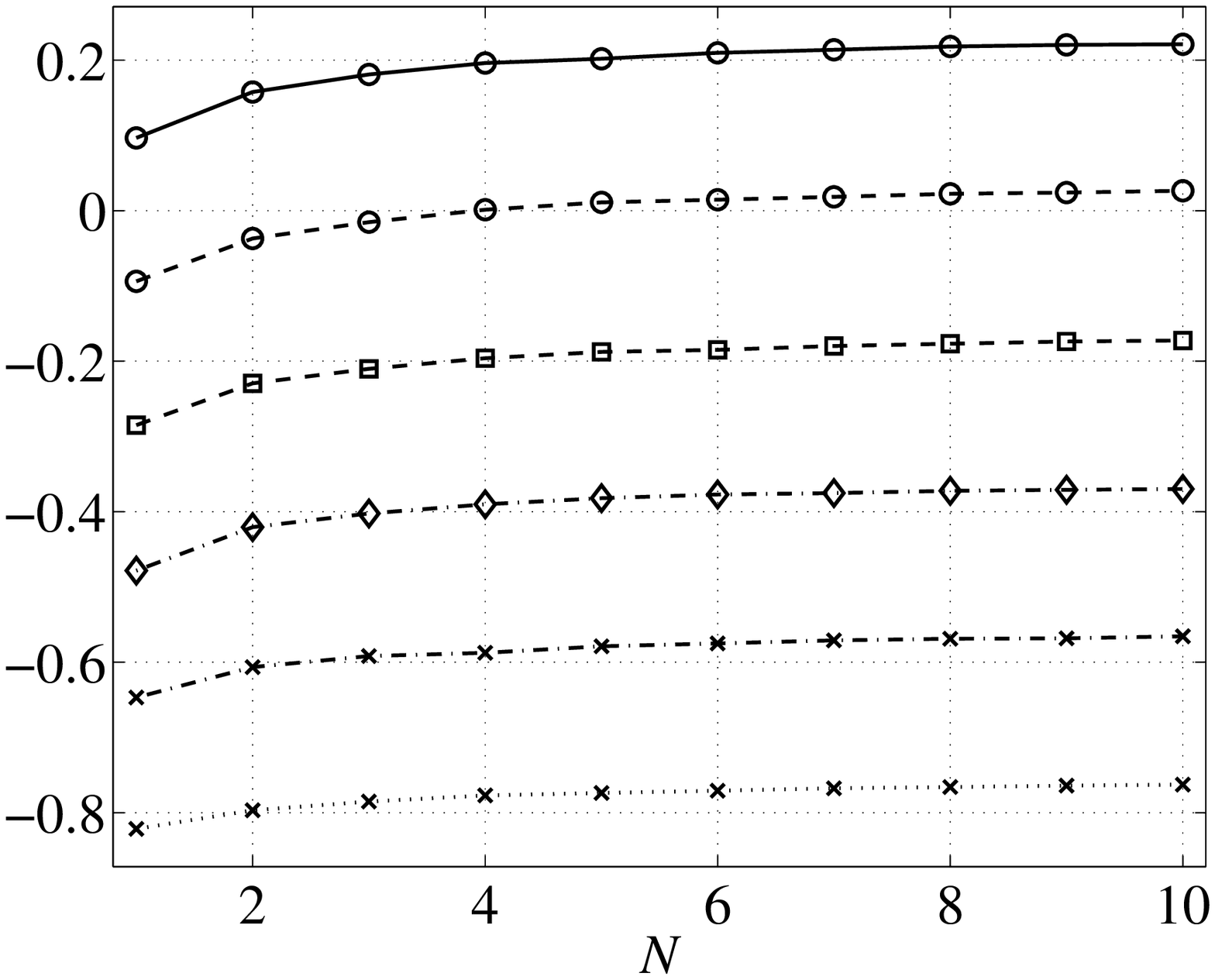} & \includegraphics[width=0.250\textwidth]{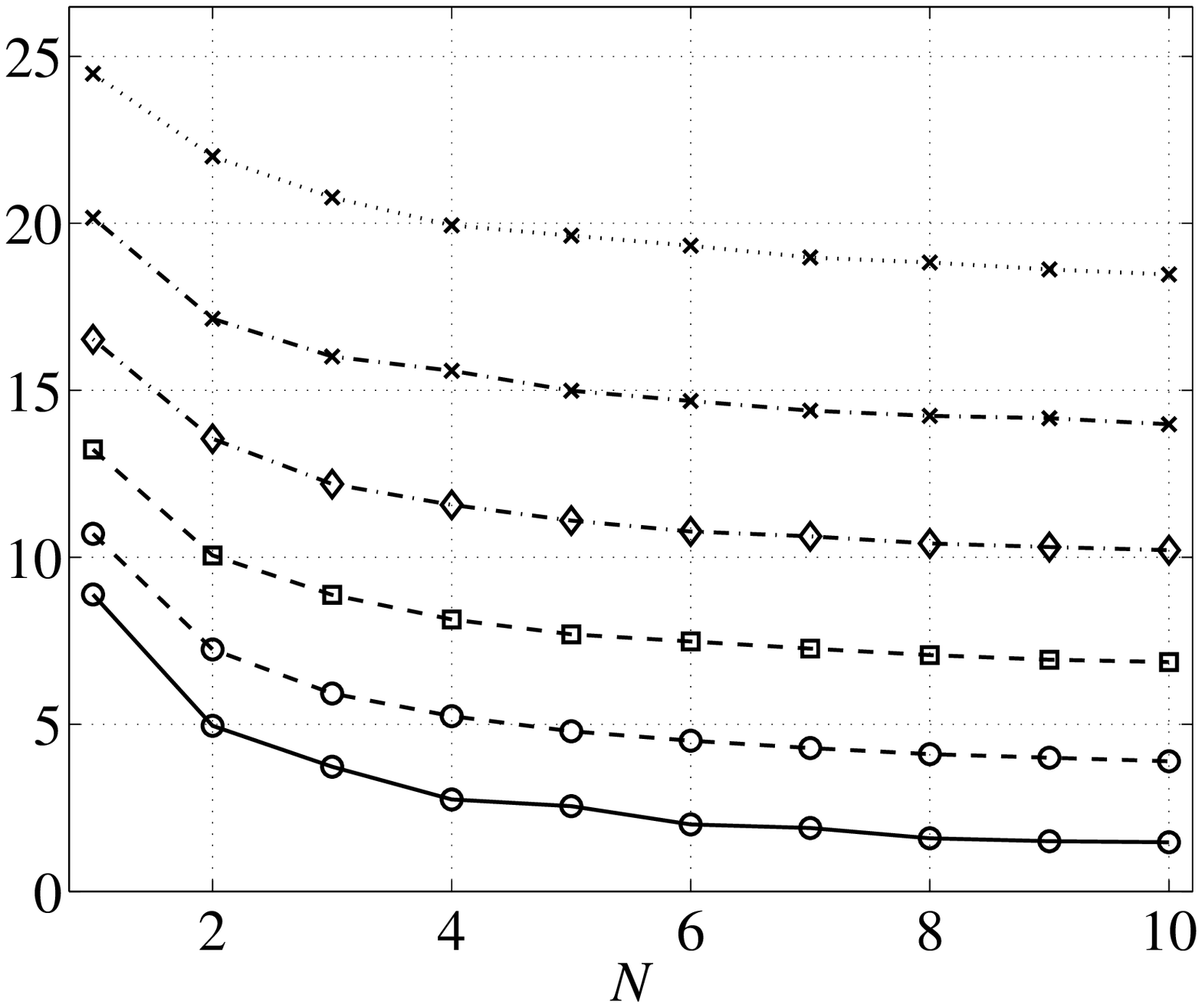} & \includegraphics[width=0.250\textwidth]{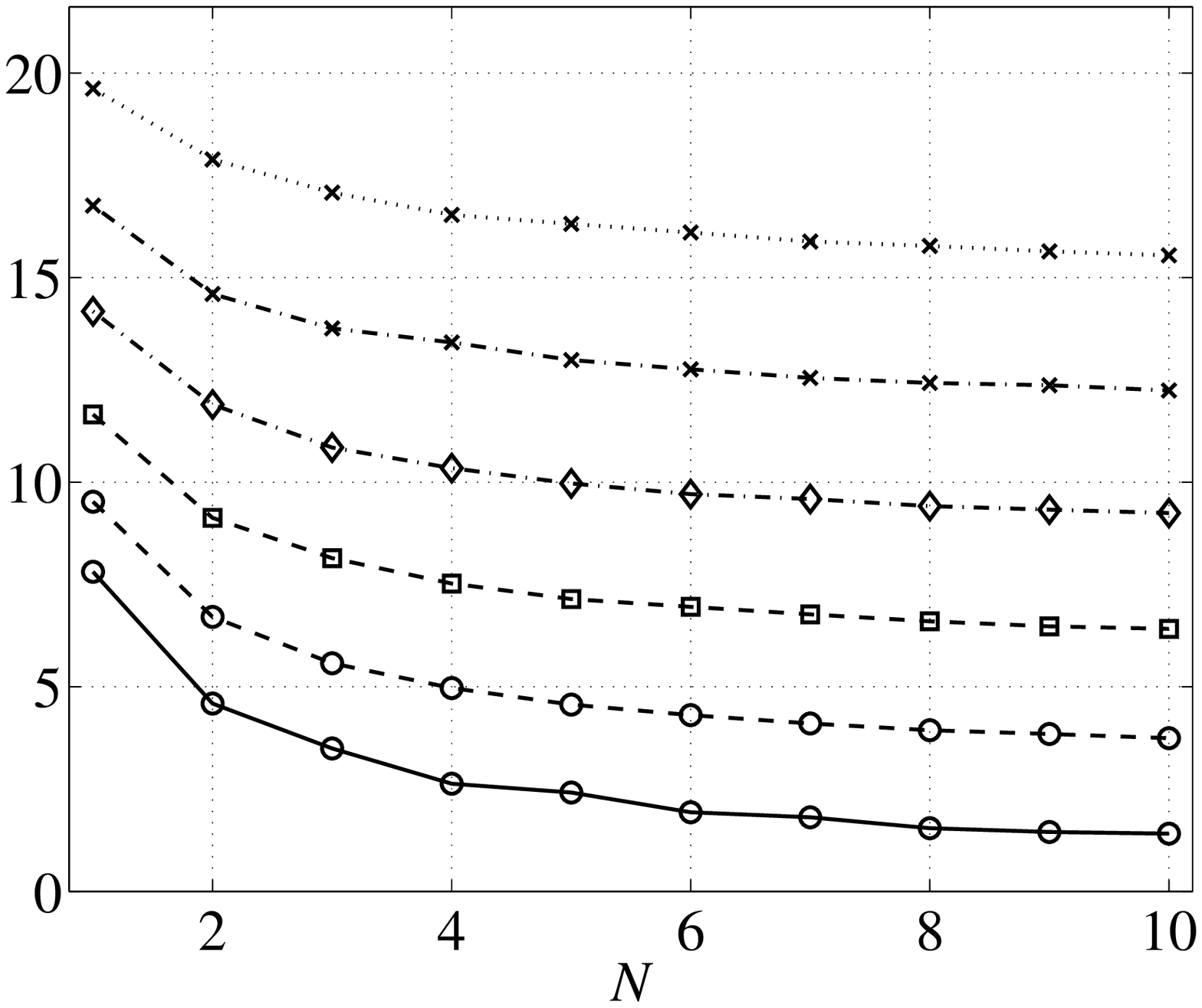} \\ 
{\scriptsize Distinguishability $\mathcal{D}$ [\%]} & {\scriptsize Bell factor $\mathcal{B}$}& {\scriptsize Mean min-to-max ratio $\mathcal{R}$ [\%]} & {\scriptsize Discrimination error $\mathcal{E}$ [\%]}   
\end{tabular}
\caption{The figures of merit as a function of $N$ using a coherent displacer $\ADis$ and an APD $\PiAPD$ with different quantum efficiencies $\eta$. The pool of candidate states consists of the two qubits of Eq. (\ref{eq:OurQubits}).}
\label{fig:CD_APD}
\end{figure*}

If, instead of a Hadamard rotation, we use a coherent displacer $\ADis_{\tau}$, we obtain the trends in Fig. \ref{fig:CD_APD}. For being a realistic device, the coherent displacer---unlike $\AHad_{\tau}$---cannot achieve a perfect similarity transformation to the ideal POVM. Indeed, none of the figures of merit are ideal for the direct measurement $N = 1$ even if we do see a violation of Bell's inequality for $\eta = 100\%$. (Bell tests with a combination of $\ADis_{\tau}$ and $\PiAPD_{\mu}$ were discussed at length in Refs. \cite{Laghaout2010, Laghaout2011, Chaves2011}.) By implementing generalized measurements, all figures of merit improve. This is particularly apparent for the discrimination error for which a generalized measurement with $N = 4$ and $\eta = 80\%$ is even slightly better than a direct measurement with ideal quantum efficiency. To re-phrase this surprising result with a classical metaphor, it is as if four myopic eyes could discern different objects as well---if not better---than a single eye of perfect vision.

\subsubsection{Homodyne detection}

\begin{figure*}
\begin{tabular}{ c c c c}
\multicolumn{4}{c}{\includegraphics[width=.45\textwidth]{legend.pdf}} \\
\includegraphics[width=0.250\textwidth]{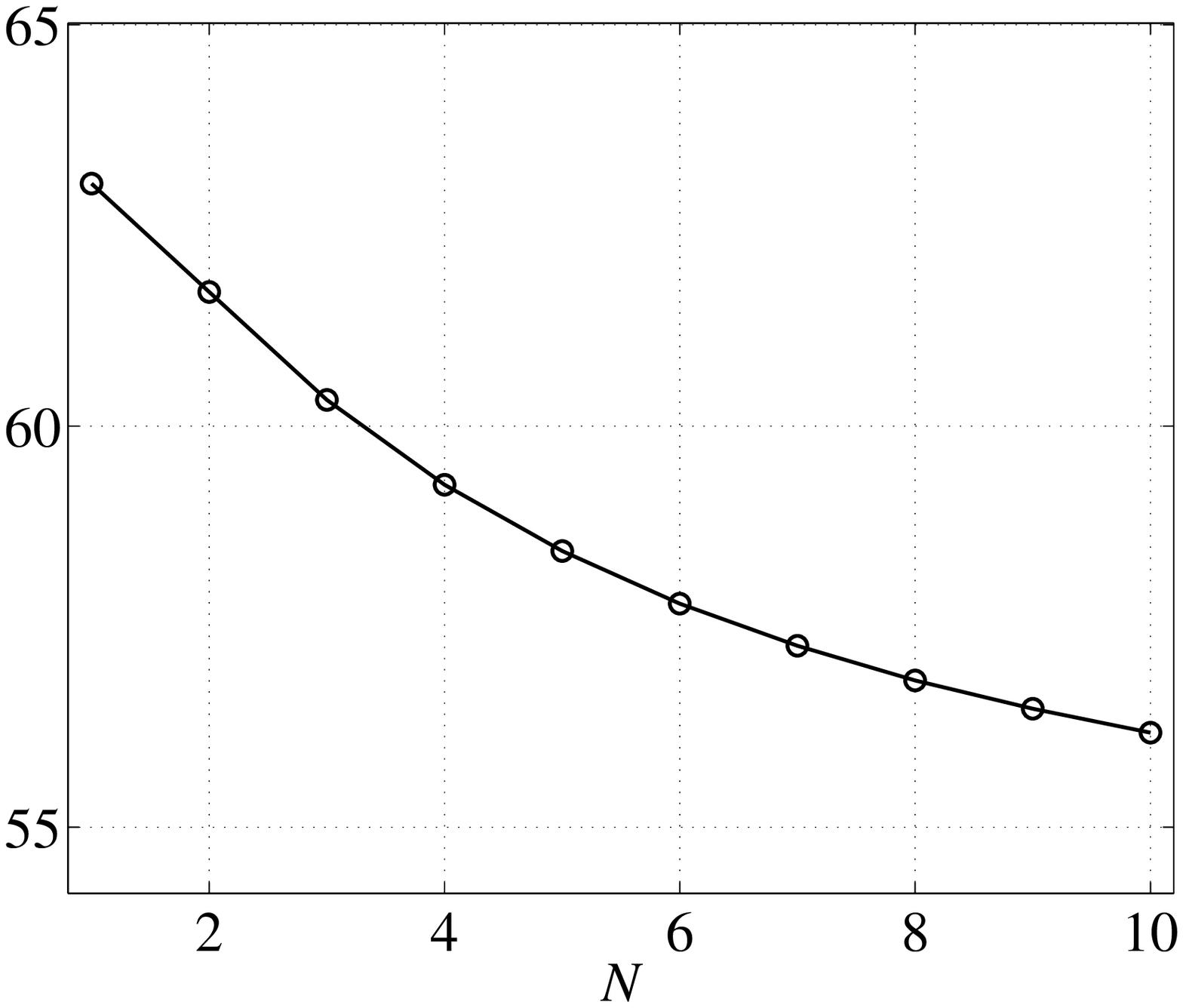} & 				\includegraphics[width=0.250\textwidth]{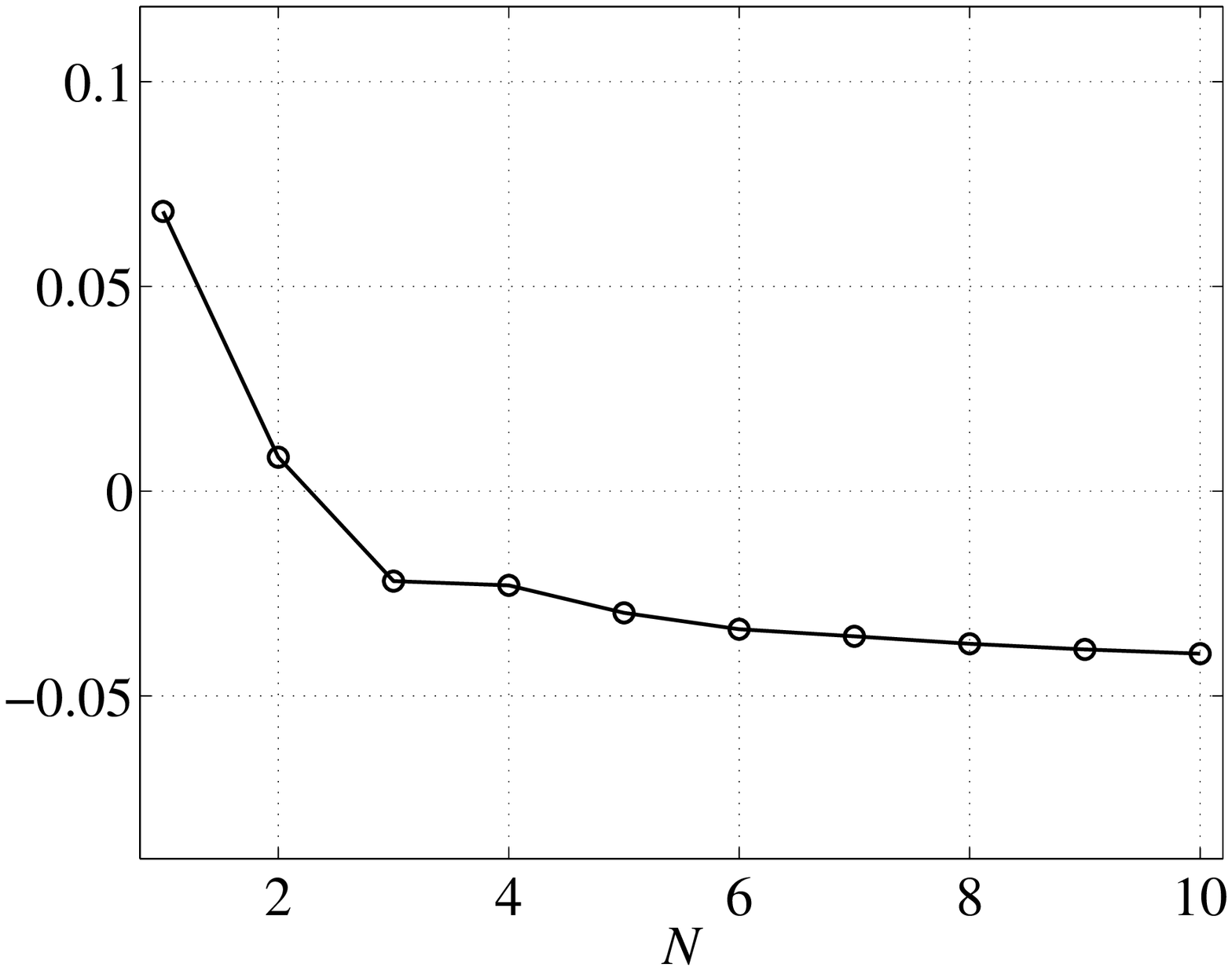} & \includegraphics[width=0.250\textwidth]{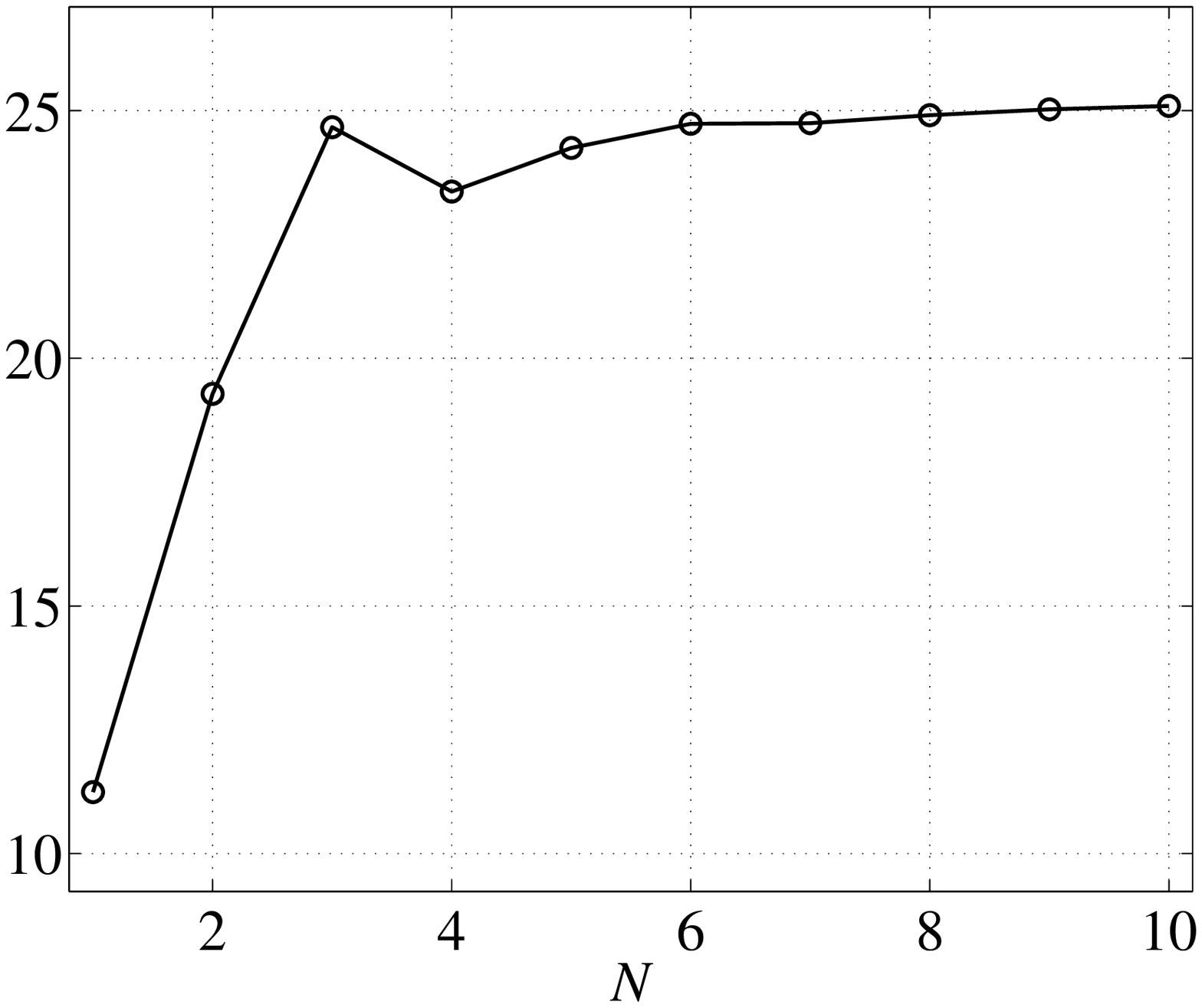} & \includegraphics[width=0.250\textwidth]{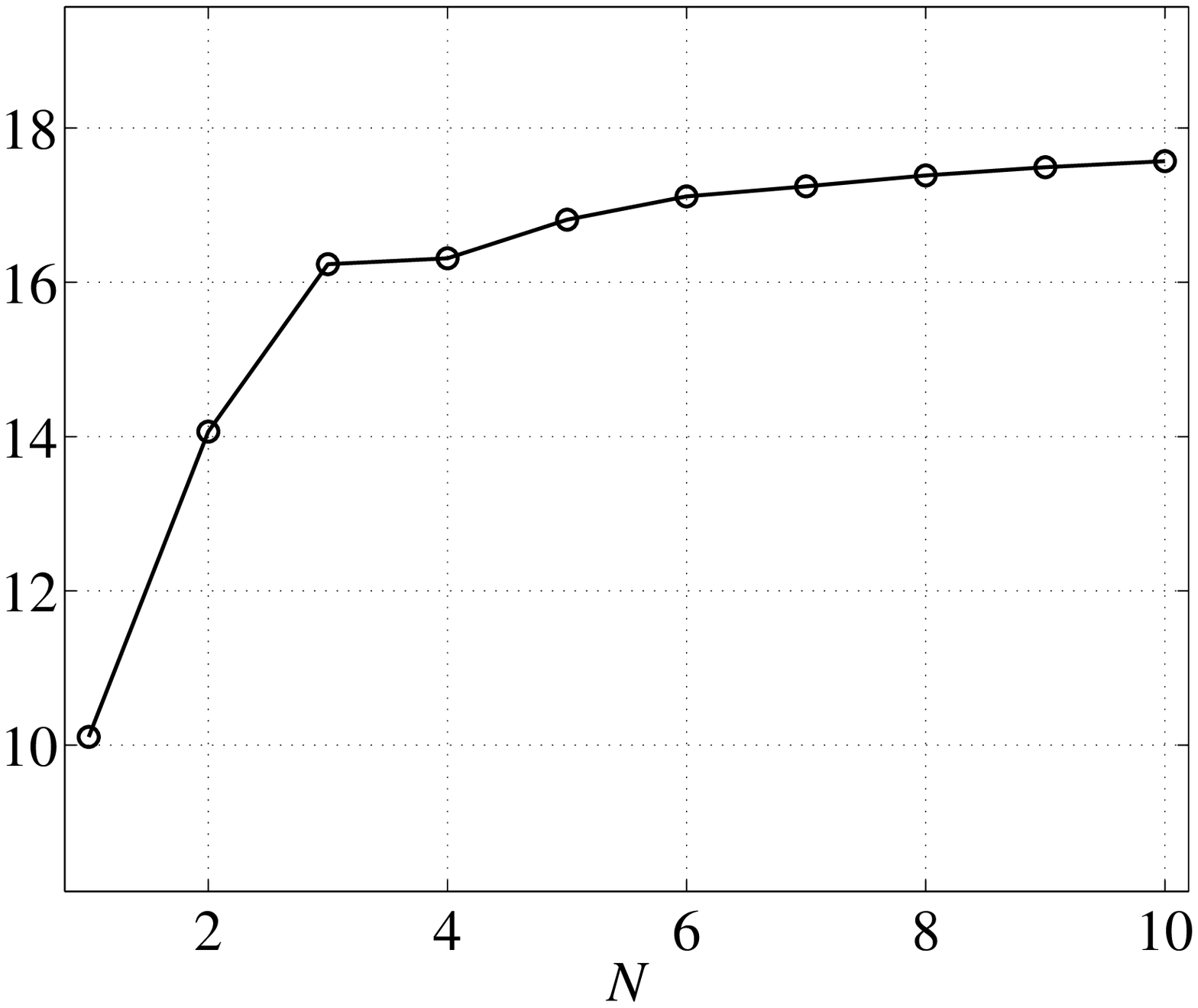} \\ 
{\scriptsize Distinguishability $\mathcal{D}$ [\%]} & {\scriptsize Bell factor $\mathcal{B}$}& {\scriptsize Mean min-to-max ratio $\mathcal{R}$ [\%]} & {\scriptsize Discrimination error $\mathcal{E}$ [\%]}   
\end{tabular}
\caption{The figures of merit as a function of $N$ using a Hadamard rotator $\AHad$ and a homodyne detector $\PiHD$ of unit quantum efficiency. The pool of candidate states consists of the two qubits of Eq. (\ref{eq:OurQubits}).}
\label{fig:HG_HD}
\end{figure*}

\begin{figure*}
\begin{tabular}{ c c c c}
\multicolumn{4}{c}{\includegraphics[width=.45\textwidth]{legend.pdf}} \\
\includegraphics[width=0.250\textwidth]{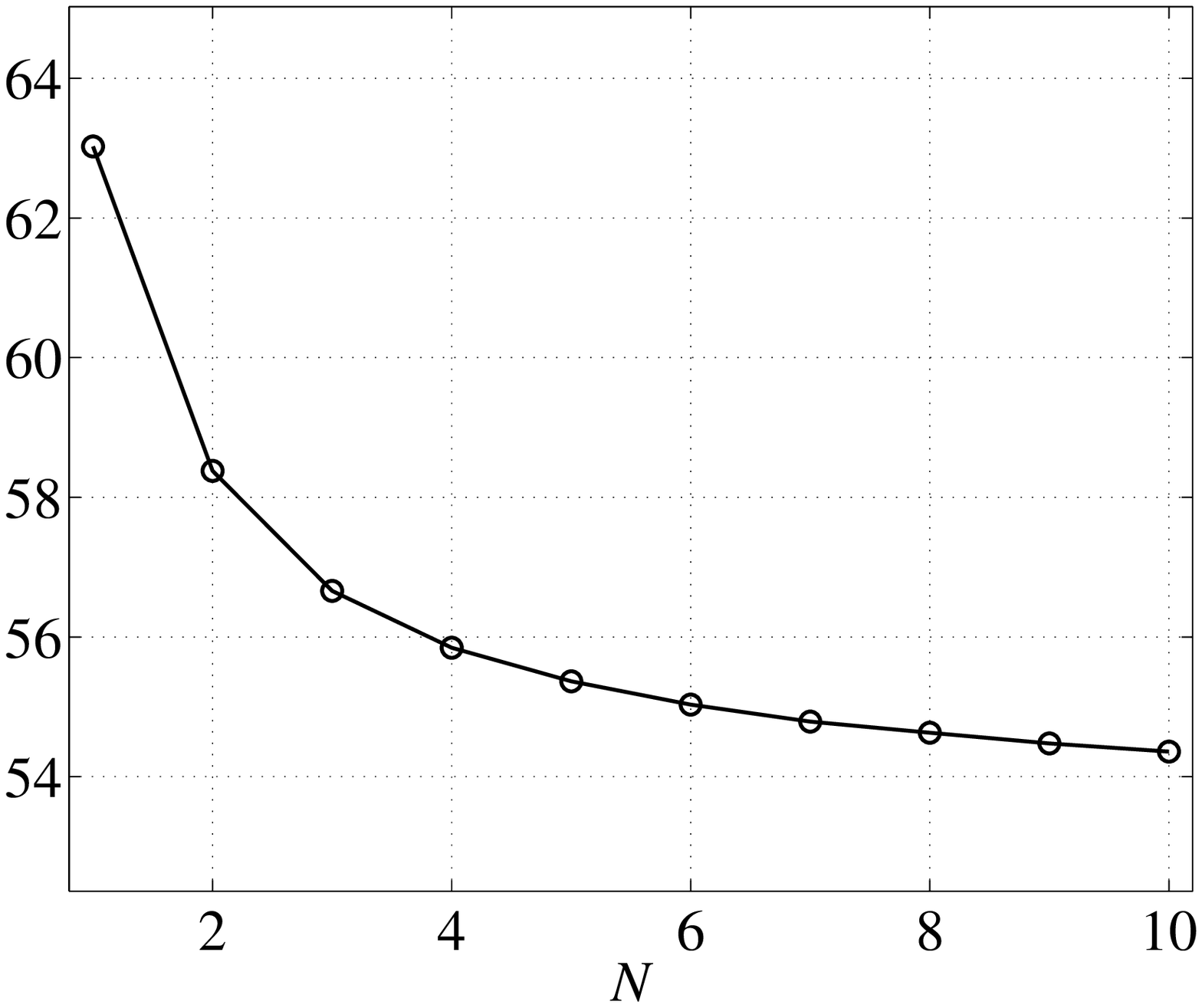} & 				\includegraphics[width=0.250\textwidth]{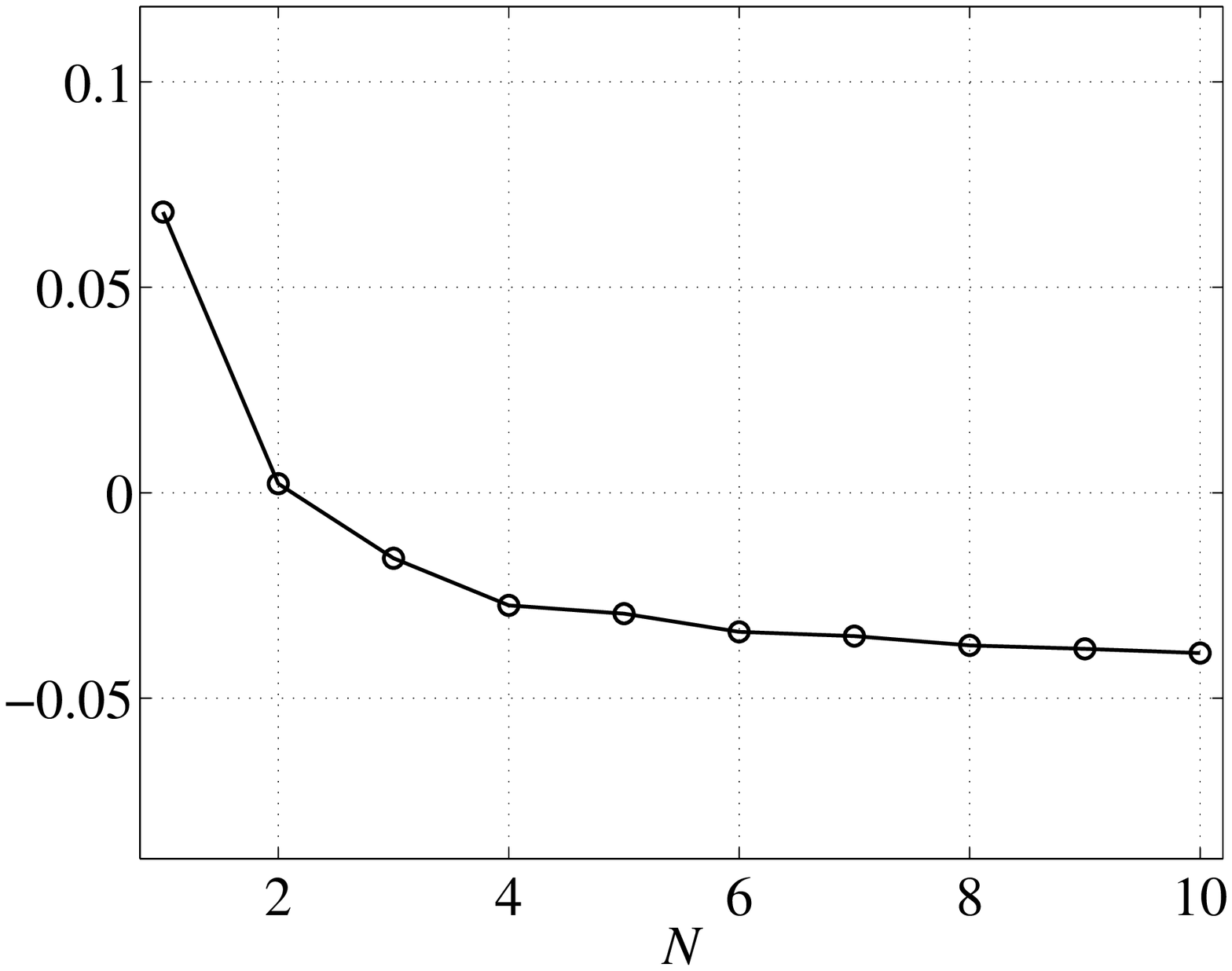} & \includegraphics[width=0.250\textwidth]{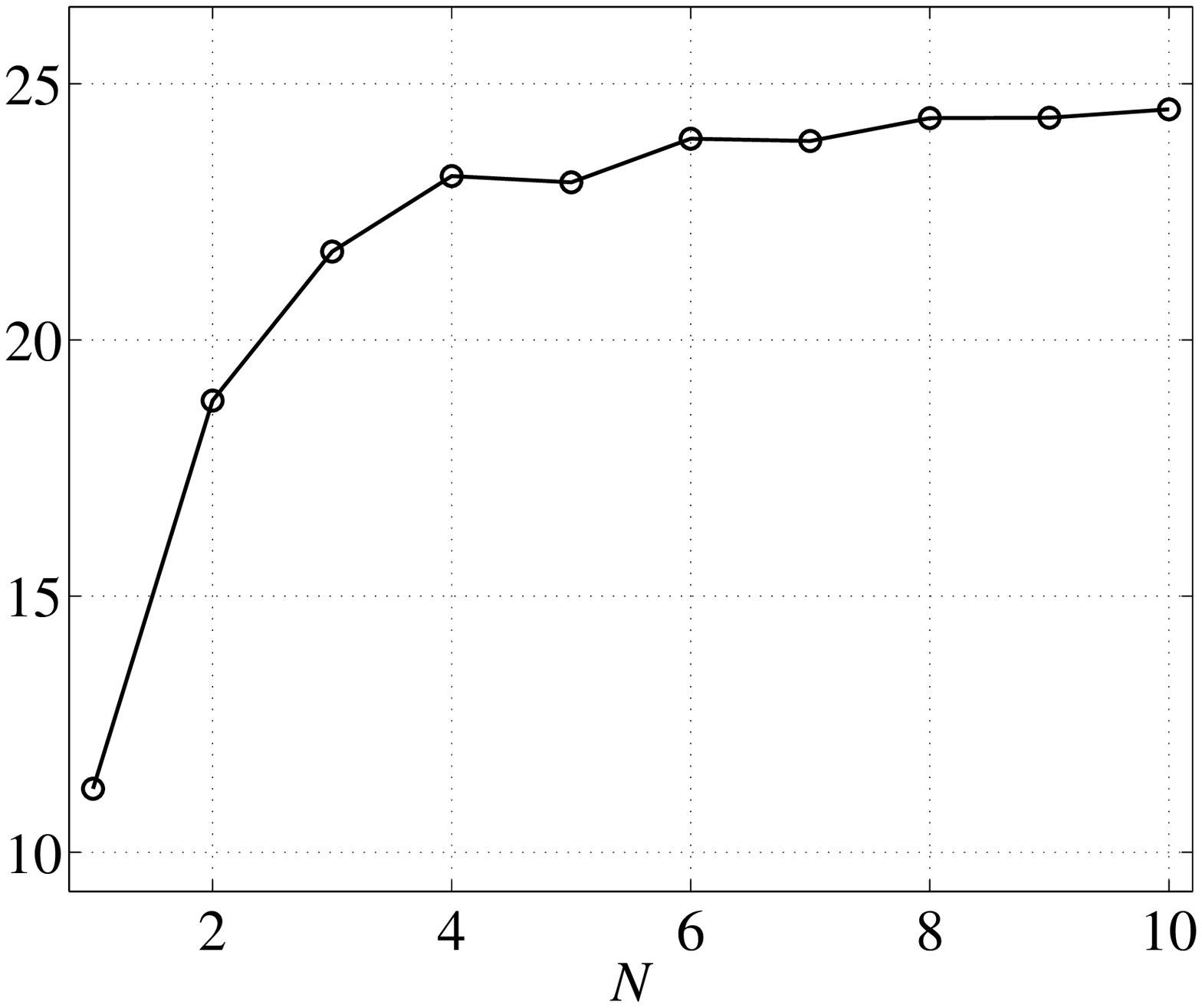} & \includegraphics[width=0.250\textwidth]{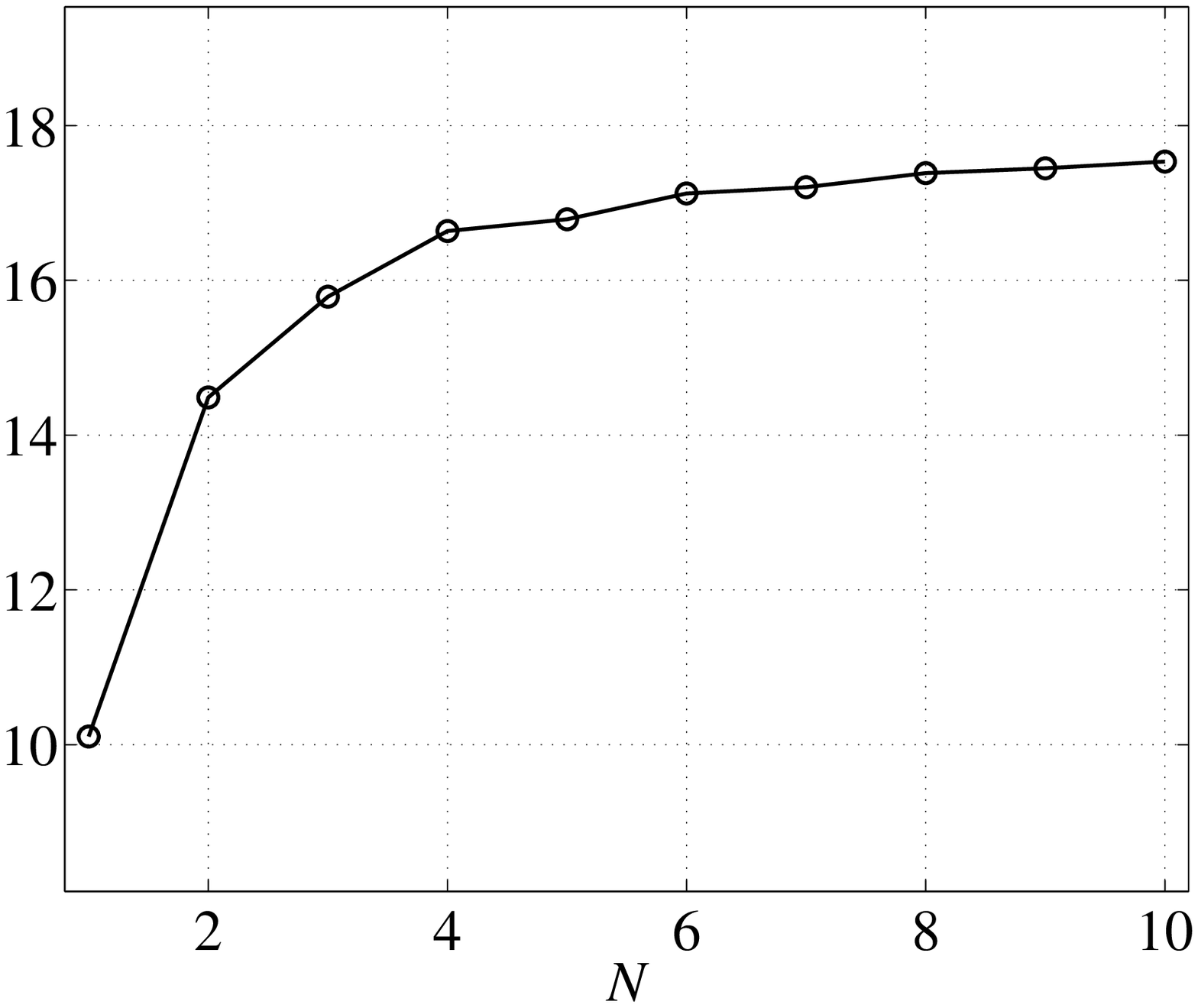} \\ 
{\scriptsize Distinguishability $\mathcal{D}$ [\%]} & {\scriptsize Bell factor $\mathcal{B}$}& {\scriptsize Mean min-to-max ratio $\mathcal{R}$ [\%]} & {\scriptsize Discrimination error $\mathcal{E}$ [\%]}   
\end{tabular}
\caption{The figures of merit as a function of $N$ using a coherent displacer $\ADis$ and a homodyne detector $\PiHD$  of unit quantum efficiency. The pool of candidate states consists of the two qubits of Eq. (\ref{eq:OurQubits}).}
\label{fig:CD_HD}
\end{figure*}

Our simulations with a homodyne detector $\PiHD_{\mu}$ clearly show that there is nothing to be gained by generalized measurements and that, on the contrary, all figures of merit suffer a degradation with higher $N$. These trends are shown in Figs. \ref{fig:HG_HD} and \ref{fig:CD_HD}. This, however, may simply mean that that the quadrature binning of Eqs. (\ref{eq:PiHD1}) and (\ref{eq:PiHD2}) is not adapted to the problem or that conjugate quadratures also need to be probed.

\subsubsection{Distinguishability as a function of $M$ and $C$}

\begin{figure*}
\begin{tabular}{ c c c c}
\multicolumn{4}{c}{\includegraphics[width=.45\textwidth]{legend.pdf}} \\
\includegraphics[width=0.250\textwidth]{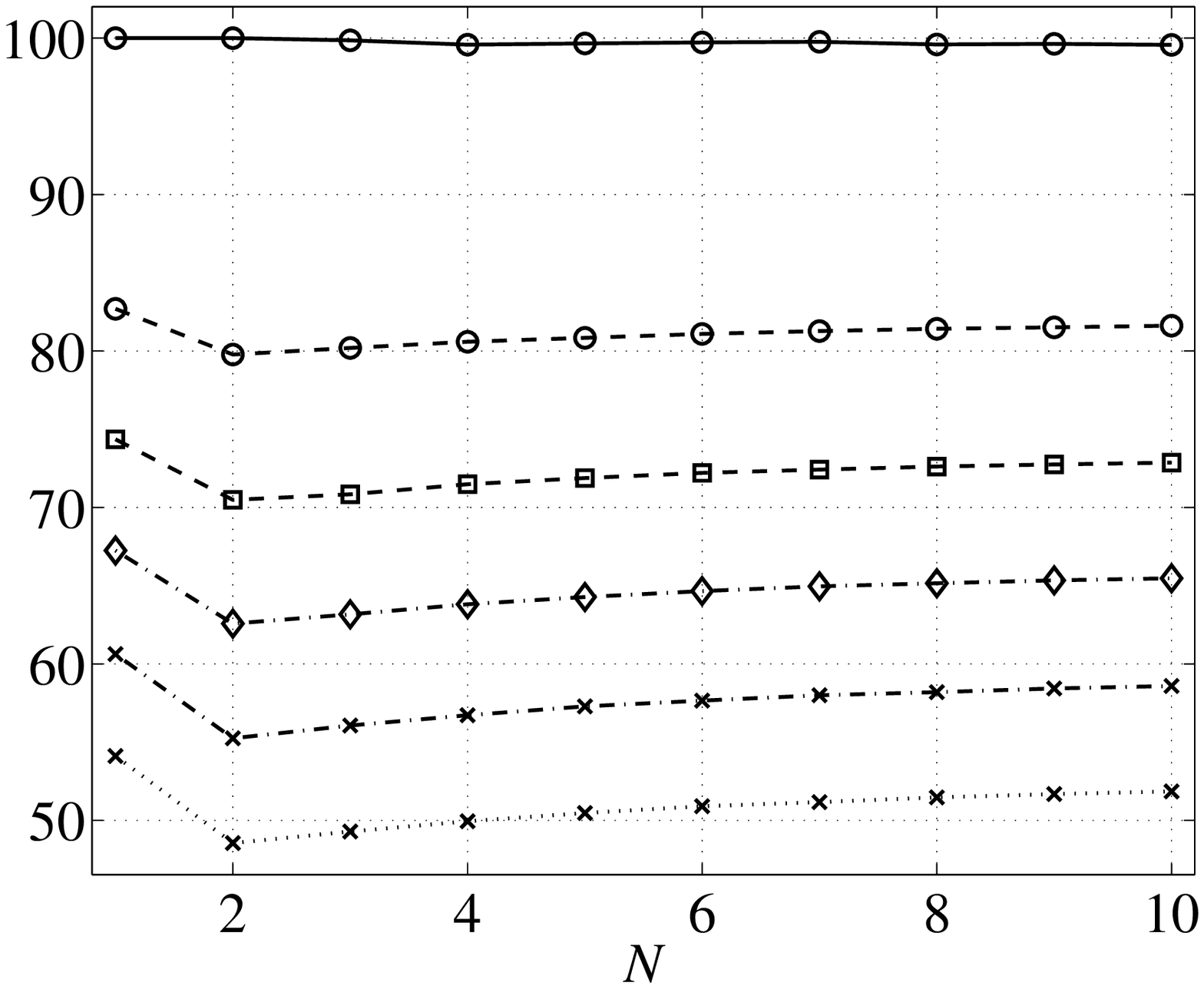} & 				\includegraphics[width=0.250\textwidth]{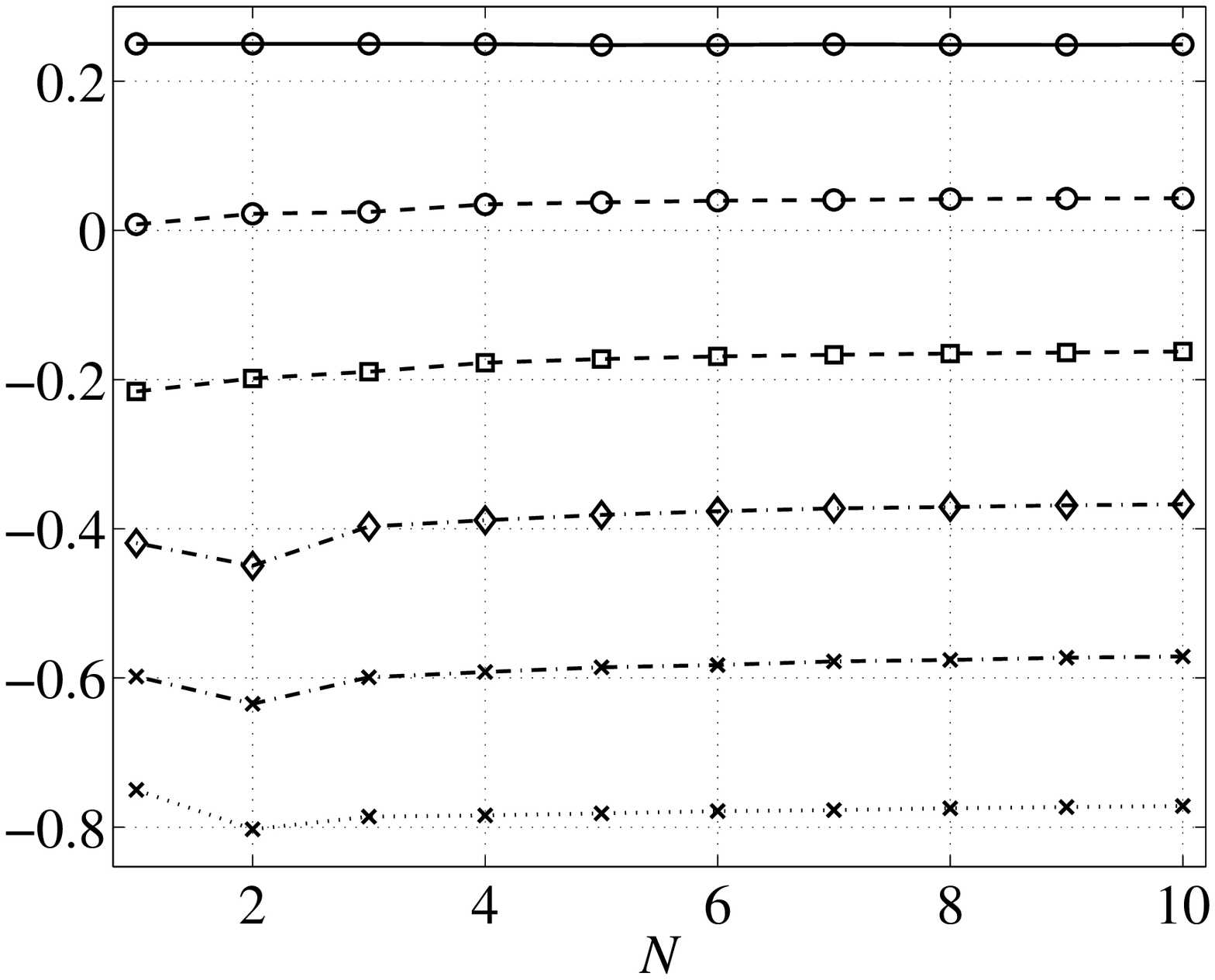} & \includegraphics[width=0.250\textwidth]{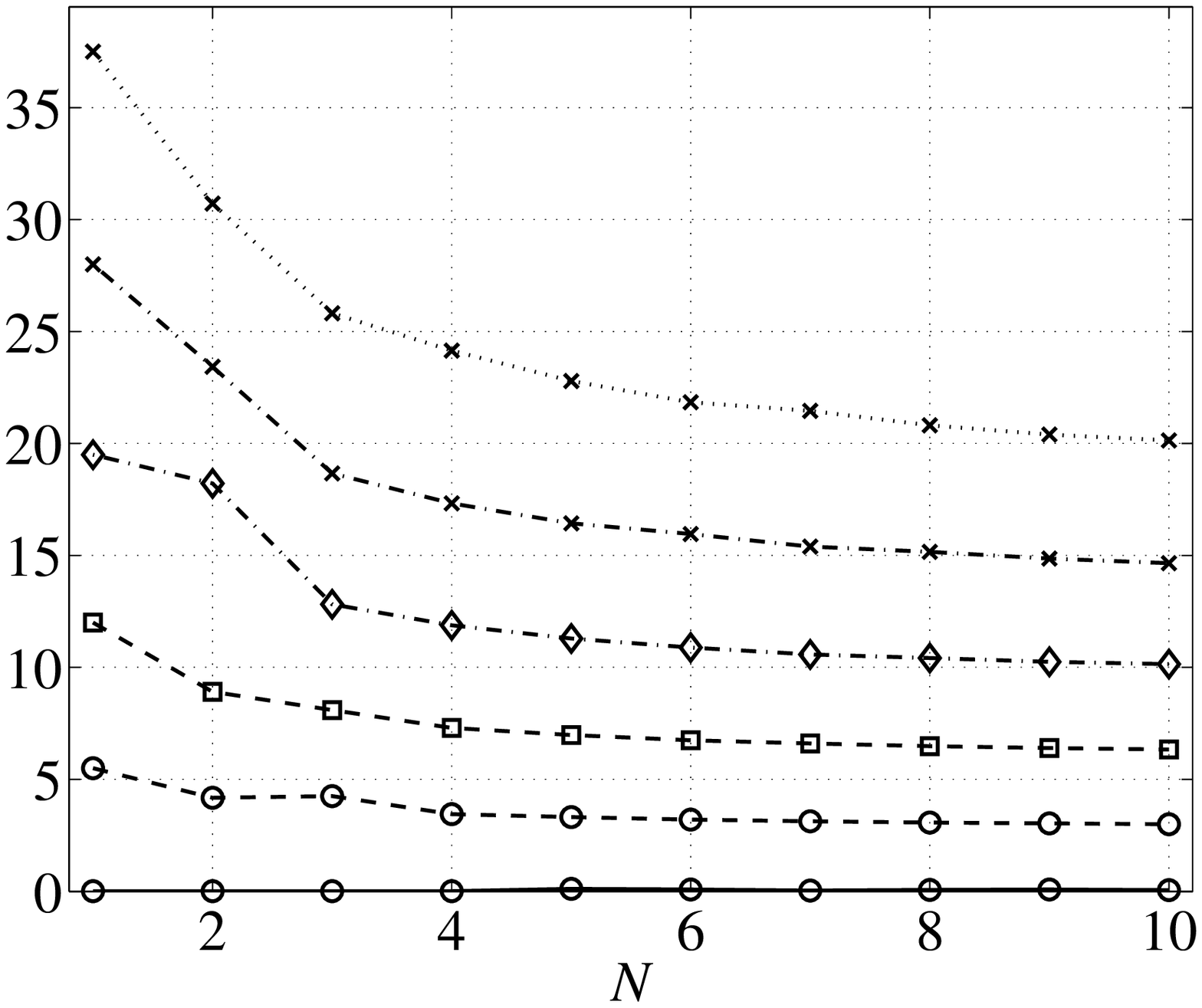} & \includegraphics[width=0.250\textwidth]{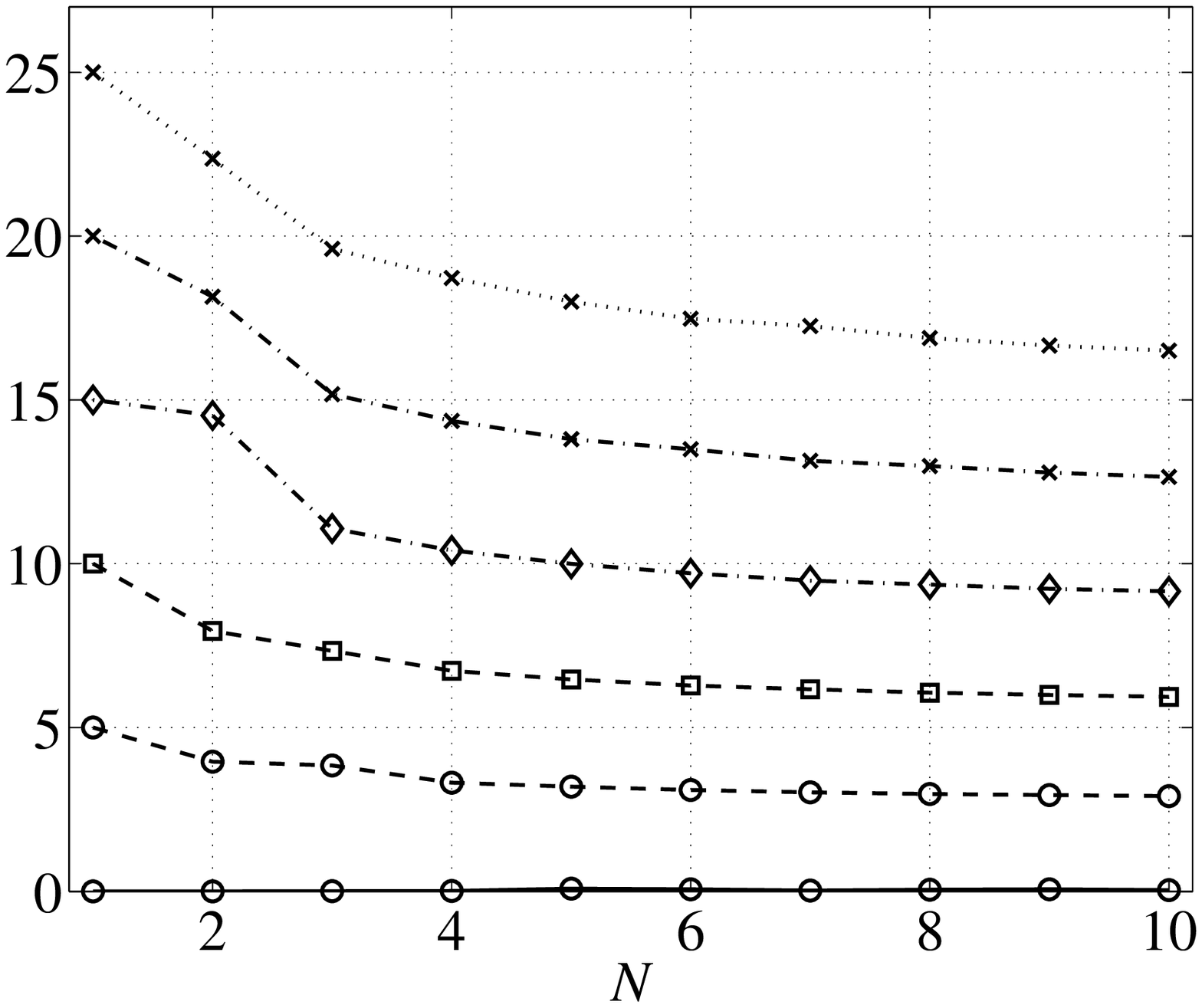} \\ 
{\scriptsize Distinguishability $\mathcal{D}$ [\%]} & {\scriptsize Bell factor $\mathcal{B}$}& {\scriptsize Mean min-to-max ratio $\mathcal{R}$ [\%]} & {\scriptsize Discrimination error $\mathcal{E}$ [\%]}   
\end{tabular}
\caption{The figures of merit as a function of $N$ using a Hadamard rotator $\AHad$ and a PNRD $\PiPNRD$ with different quantum efficiencies $\eta$. The PNRD has three possible outcomes $M = 3$ and saturates at $\nsat = 2$ photon counts. The pool of candidate states consists of the two qubits of Eq. (\ref{eq:OurQubits}).}
\label{fig:HG_PNRD}
\end{figure*}

\begin{figure*}
\begin{tabular}{ c c c c}
\multicolumn{4}{c}{\includegraphics[width=.45\textwidth]{legend.pdf}} \\
\includegraphics[width=0.250\textwidth]{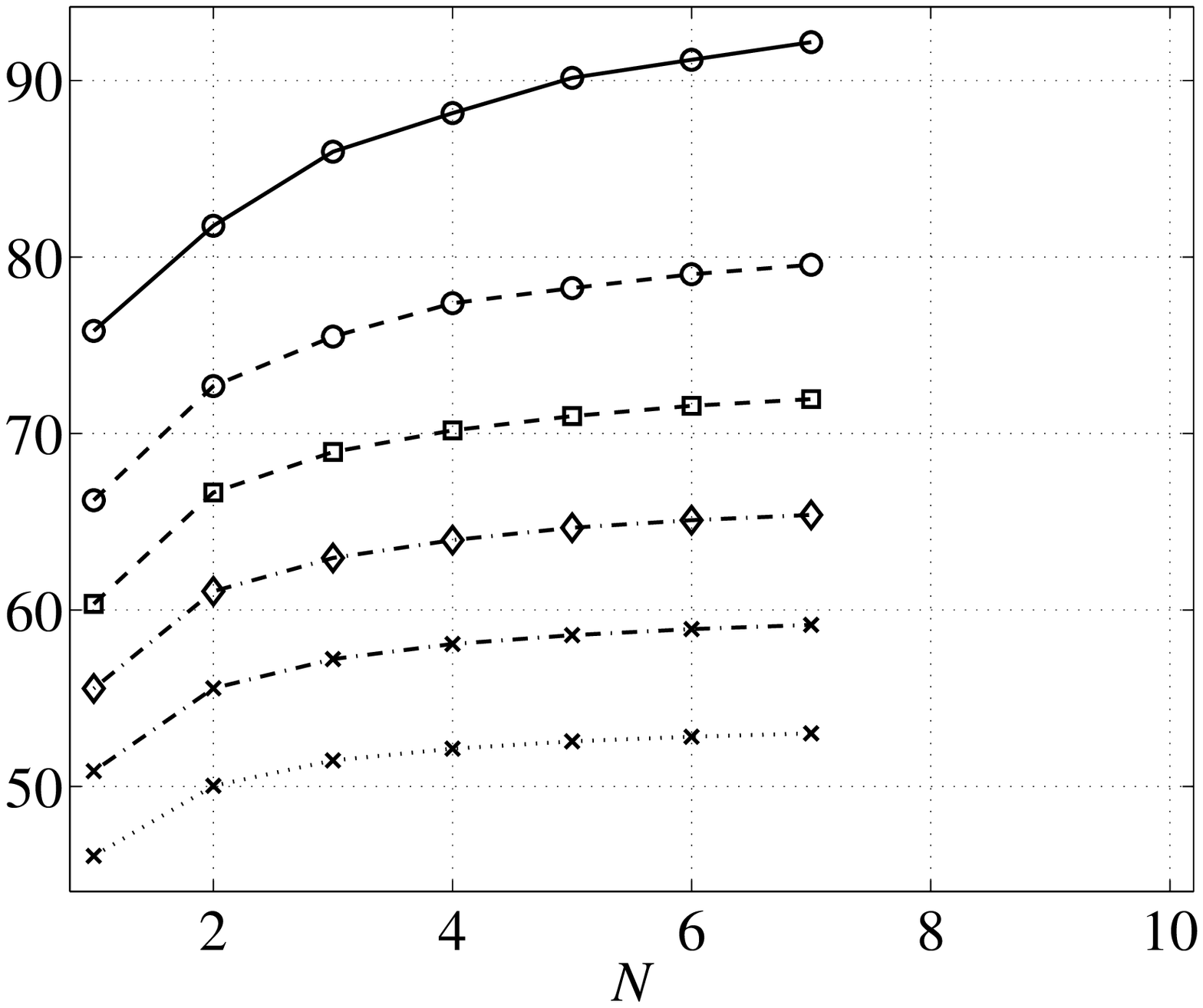} & 				\includegraphics[width=0.250\textwidth]{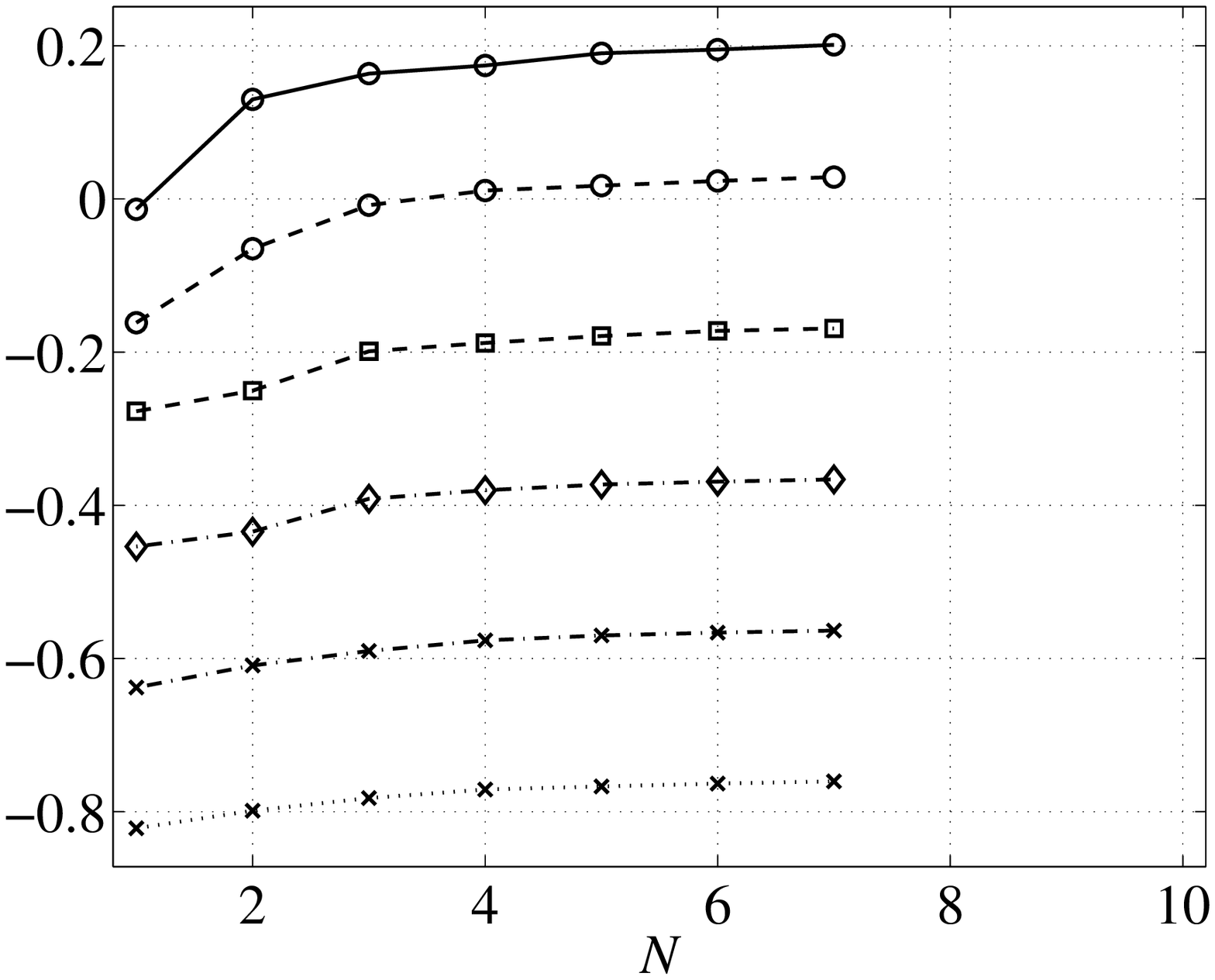} & \includegraphics[width=0.250\textwidth]{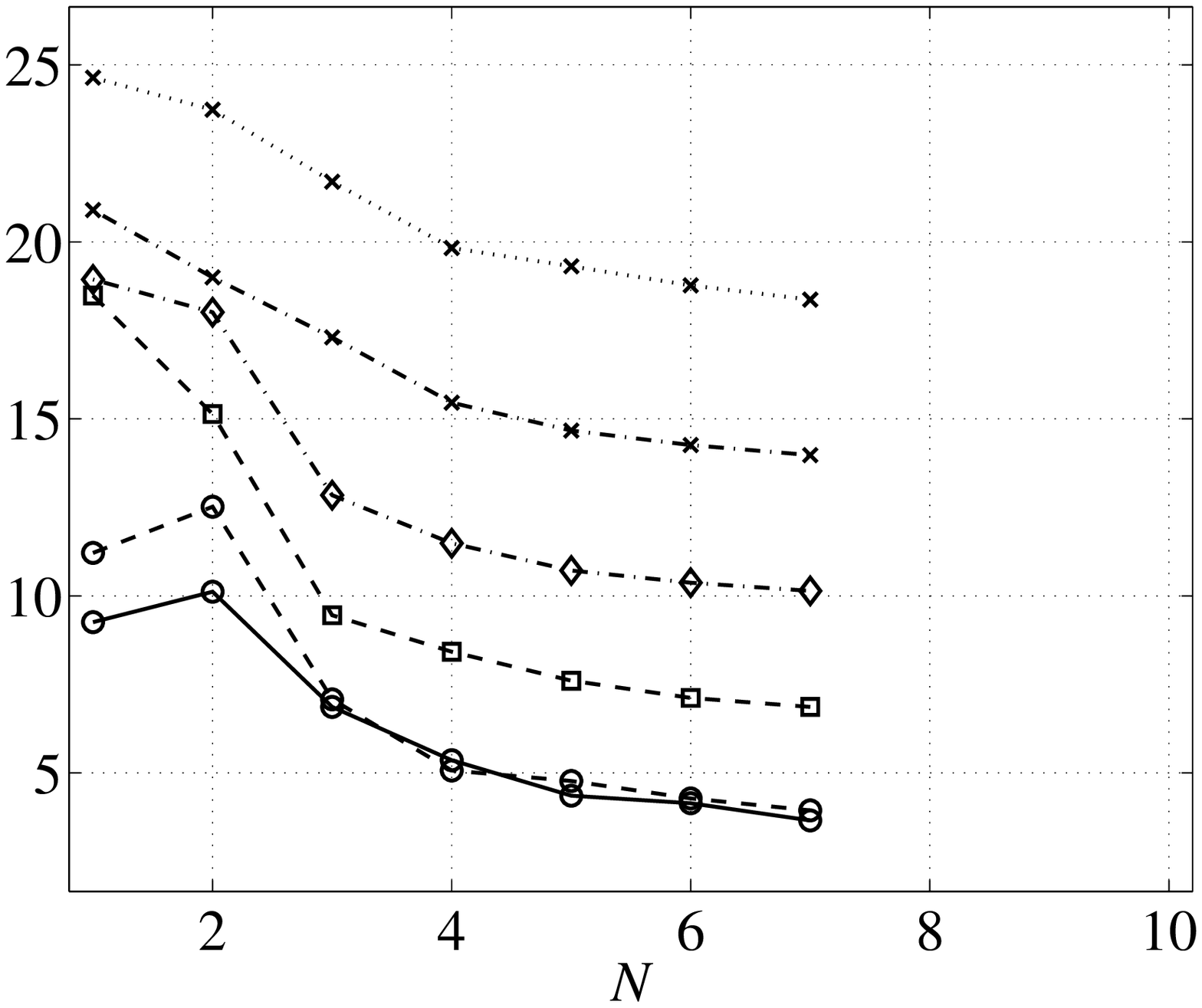} & \includegraphics[width=0.250\textwidth]{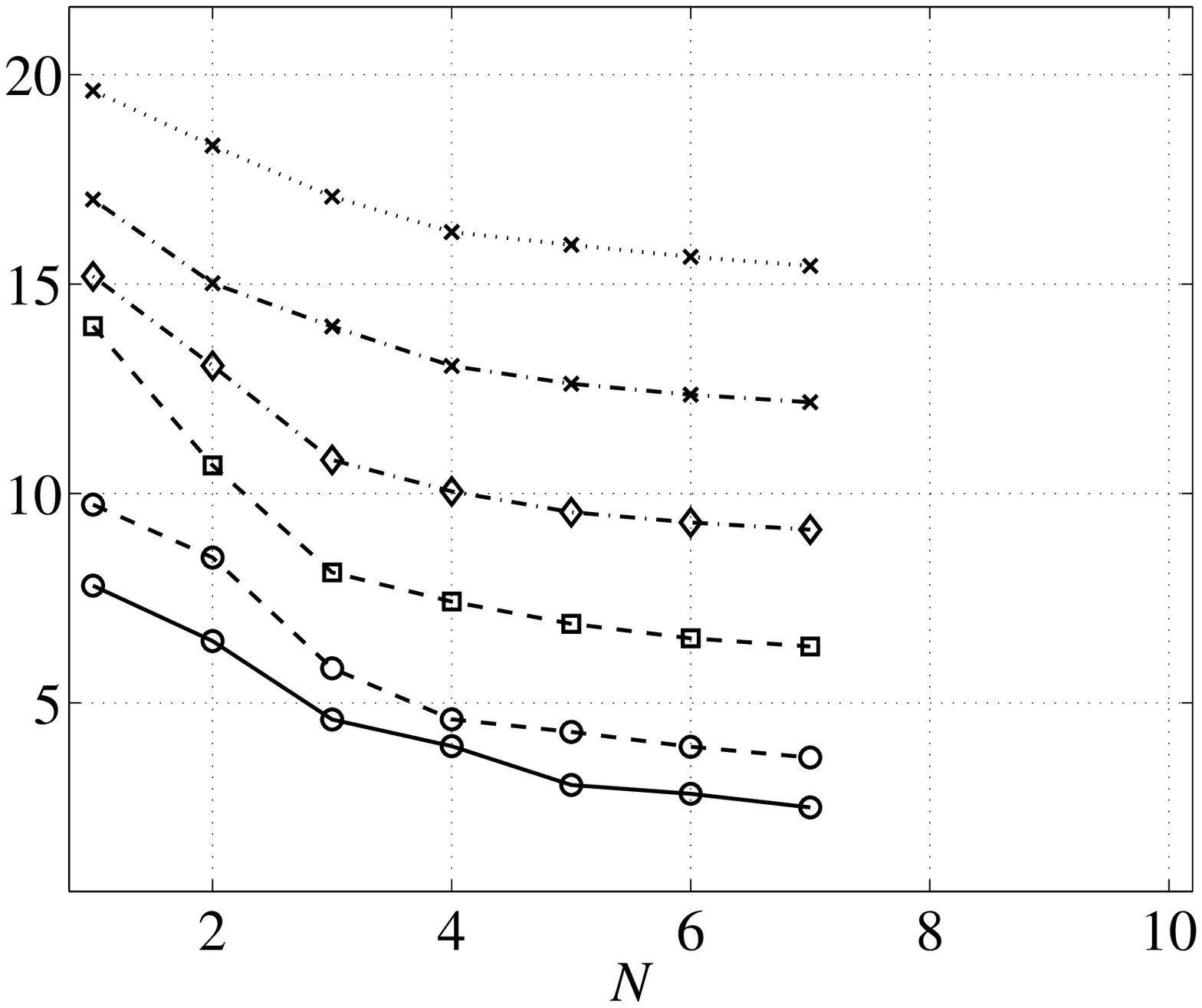} \\ 
{\scriptsize Distinguishability $\mathcal{D}$ [\%]} & {\scriptsize Bell factor $\mathcal{B}$}& {\scriptsize Mean min-to-max ratio $\mathcal{R}$ [\%]} & {\scriptsize Discrimination error $\mathcal{E}$ [\%]}   
\end{tabular}
\caption{The figures of merit as a function of $N$ using a coherent displacer $\ADis$ and a PNRD $\PiPNRD$ with different quantum efficiencies $\eta$. The PNRD has three possible outcomes $M = 3$ and saturates at $\nsat = 2$ photon counts. The pool of candidate states consists of the two qubits of Eq. (\ref{eq:OurQubits}). The simulations took an inordinate amount of time beyond $N = 7$ and had to be aborted. This is due to the combined demand in memory resources from the coherent displacer, which occupies larger matrices in Fock space than the qubit rotator, and the higher exponential growth of the decision tree with the $M = 3$ instead of $M = 2$.}
\label{fig:CD_PNRD}
\end{figure*}

\begin{figure}
  \centering\includegraphics[width=.9\columnwidth]{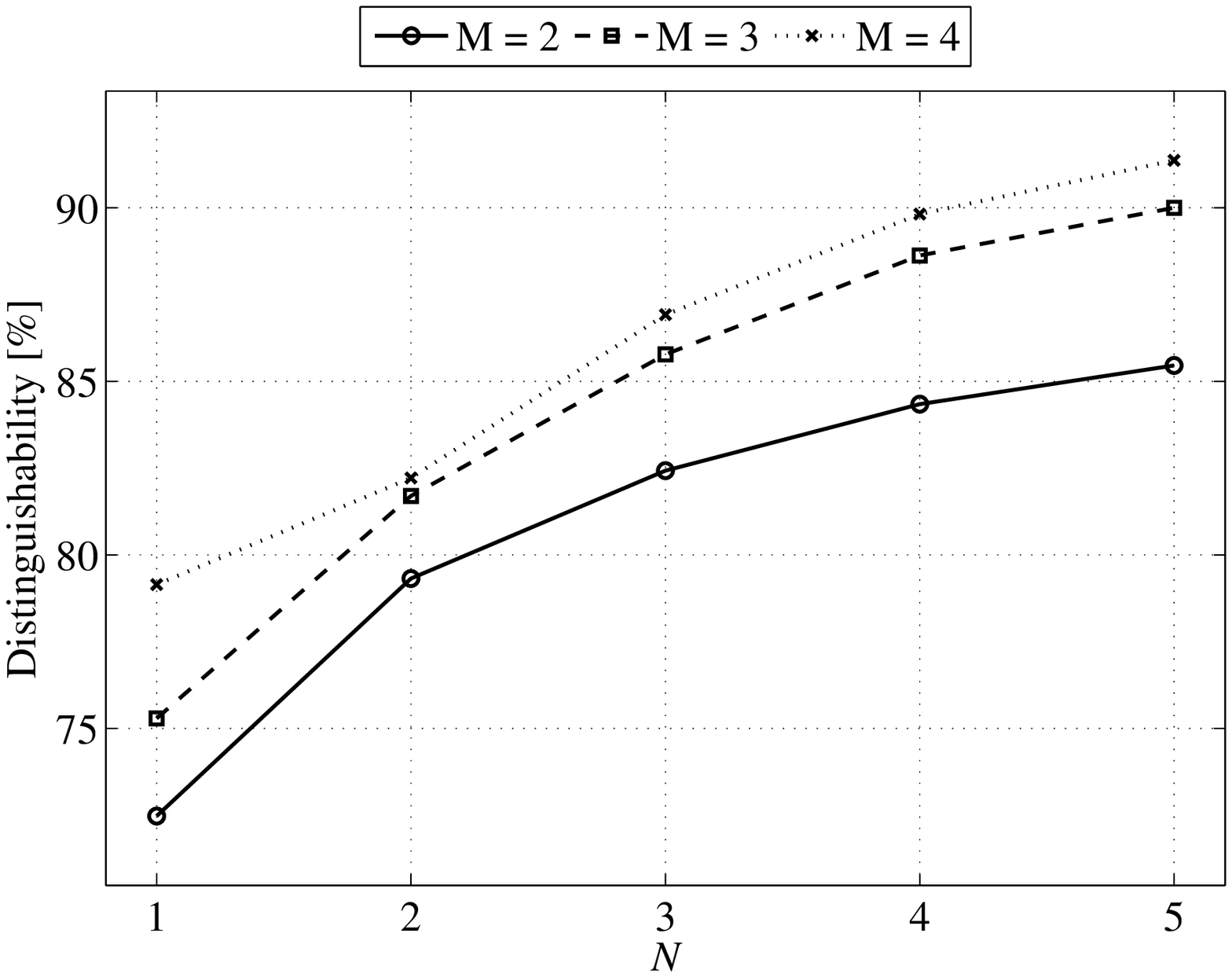}
  \caption{Dependence of the distinguishability $\mathcal{D}$ on the number $M$ of measurement outcomes at each node. Each measurement consists of a coherent displacer $\ADis$ and a PNRD $\PiPNRD$ which saturates at 1, 2, and 3 photons, respectively. The pool of candidate states consisted of the two qubits of Eq. (\ref{eq:OurQubits}). In order to bring up any differences in photon counts which could be revealed by a higher photon count, we have extended the parameter range for the displacement to  $\mathcal{T} = \hak{-2, 2}$ instead of the shorter segment $\hak{-1, 1}$ used so far. We can clearly see that the distinguishability improves with $M$ although the same cannot be said of the other figures of merit in this particular example (cf. Fig. \ref{fig:CD_PNRD}).}
	\label{fig:M_dependence_CD_PNRD}
\end{figure}

\begin{figure}
  \centering\includegraphics[width=.9\columnwidth]{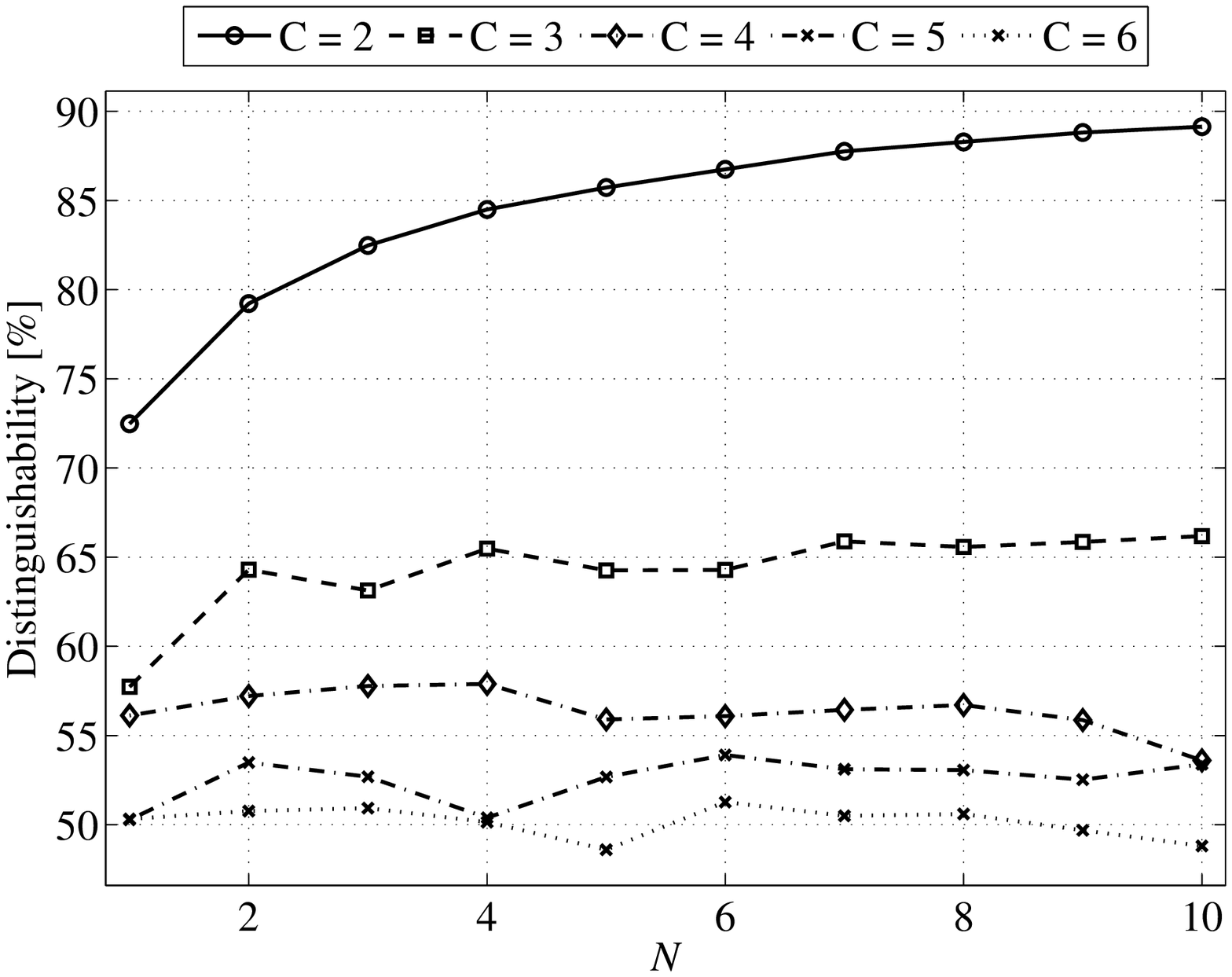}
  \caption{Dependence of the distinguishability $\mathcal{D}$ on the number $C$ of candidate qubits in the pool defined in Eq. (\ref{eq:QubitPool}). Each measurement consists of a coherent displacer $\ADis$ and an APD $\PiAPD$ of unit quantum efficiency. For this particular combination, the generalization of the measurement to $N \ge 2$ mostly benefits pools made up of two or three  qubits.}
	\label{fig:C_dependence_CD_APD}
\end{figure}

Figures \ref{fig:HG_PNRD} and \ref{fig:CD_PNRD} make use of $\AHad_{\tau}$ and $\ADis_{\tau}$, respectively, except that instead of an APD, we employ a photon number resolving detector which saturates at $\nsat = 2$. Although the overall trend is the same as that of Figs. \ref{fig:HG_APD} and \ref{fig:CD_APD}, the figures of merit do not perform as well, except for the distinguishability (Fig. \ref{fig:M_dependence_CD_PNRD}). This might be due to the fact that for only two candidate states ($C=2$), the decision tree---now made up of three-pronged nodes ($M = 3$)---scatters into too many leaves with too little statistical value. This problem is referred to as fragmentation and we shall come back to it in Sec. \ref{sec:Issues}. The response of the figures of merit to an increased output cardinality $M$ is an interesting topic in its own right but we shall not expand on it here. The same goes for the behavior of the figures of merit with respect to a larger pool of candidate states. Figure \ref{fig:C_dependence_CD_APD} displays the distinguishability for different values of $C$ as a function of $N$ using $\AHad_{\tau}$ and $\PiAPD_{\mu}$. (Recall how the candidate states are sampled from the Bloch sphere in Eq. (\ref{eq:QubitPool}).) An improvement with $N$ is mostly witnessed for $C = 2$, whereas larger pools $C \ge 4$ do not exhibit any sensible improvement with increased $N$. Here again, we should leave open the possibility that a different set of operations could improve the results even for larger $C$. (The stagnation witnessed for $C \ge 4$ cannot be solely due to the non-orthogonality of the candidate states since the plot with $C = 3$, which also features non-orthogonal states, does improve with $N$.)

\subsection{Computational considerations}
\label{sec:Issues}

We already mentioned the heavy developmental cost that is incurred by the exponential growth of the decision trees. Even with parallelized simulations, most of our calculations were too demanding to implement beyond $N \gtrsim 10$. The problem is compounded if one includes von Neumann projectors with more than two possible outcomes. For example, our runs with $\ADis_{\tau}$ and $\PiPNRD_{\mu}$ had to be aborted already at $N = 8$ (cf. Fig. \ref{fig:CD_PNRD}). It is therefore worth mentioning a few considerations that cut down or at least help us evaluate the computational costs. 

Consider for example dynamic programming, which is one of the methods that could achieve a global optimization of the Bayesian network. In this case, all possible combinations of parameters $\tau$ in the range $\mathcal{T}$ need to be tested for all internal nodes. The run time complexity is then given by
\beq
T(N) \in \mathcal{O}\tes{S_{\mathcal{T}}^{\frac{M^N-1}{M-1}}}, 
\eeq
where $S_{\mathcal{T}}$ is the number of parameter points sampled from the parameter range. If we are to opt for greedy algorithms instead, the run time complexity is significantly reduced to 
\beq
T(N) \propto S_{\mathcal{T}} \frac{M^N-1}{M-1} \in \mathcal{O}\tes{M^N},
\eeq
where we assume that the algorithm operates as a single-pass traversal of the tree. One way to get rid of the exponential dependence on $N$ is to build a dictionary of decisions which can be recycled at each node. This dictionary consists of a $C$-dimensional table of all the possible states on the entire Bloch ball that the qubits can decohere into. For any such combination of possible states, the entries of the table will store the parameter $\tau_{o}$ which optimizes the figure of merit. This technique is of course demanding in its preparation stage as the entire Bloch ball will have to be discretized into sufficiently many C-tuples of sample points. However, it represents the best way to scale the problem without the exponential cost on $N$.

\begin{figure*}
\begin{tabular}{ c c c c}
\includegraphics[width=0.250\textwidth]{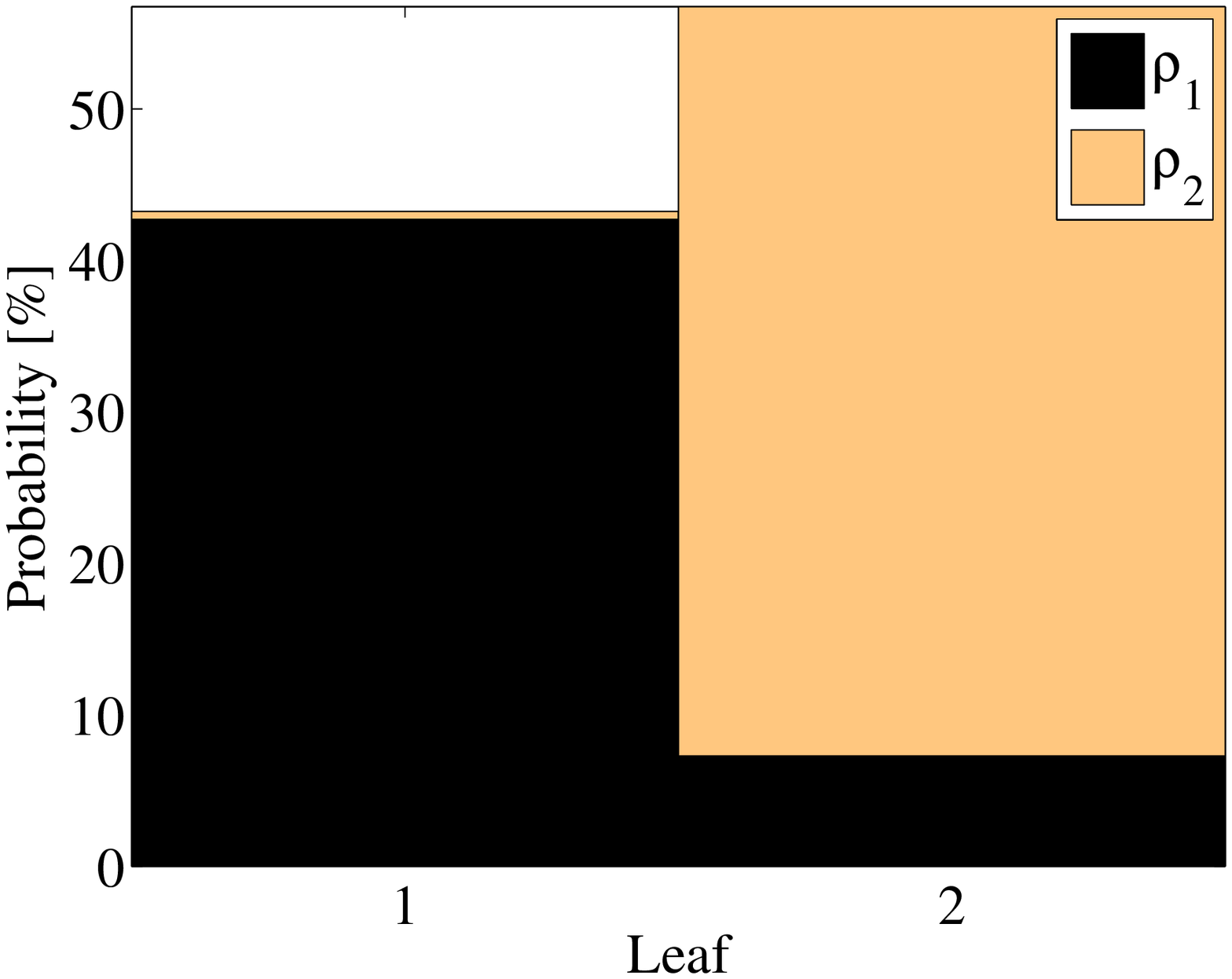} & 				\includegraphics[width=0.250\textwidth]{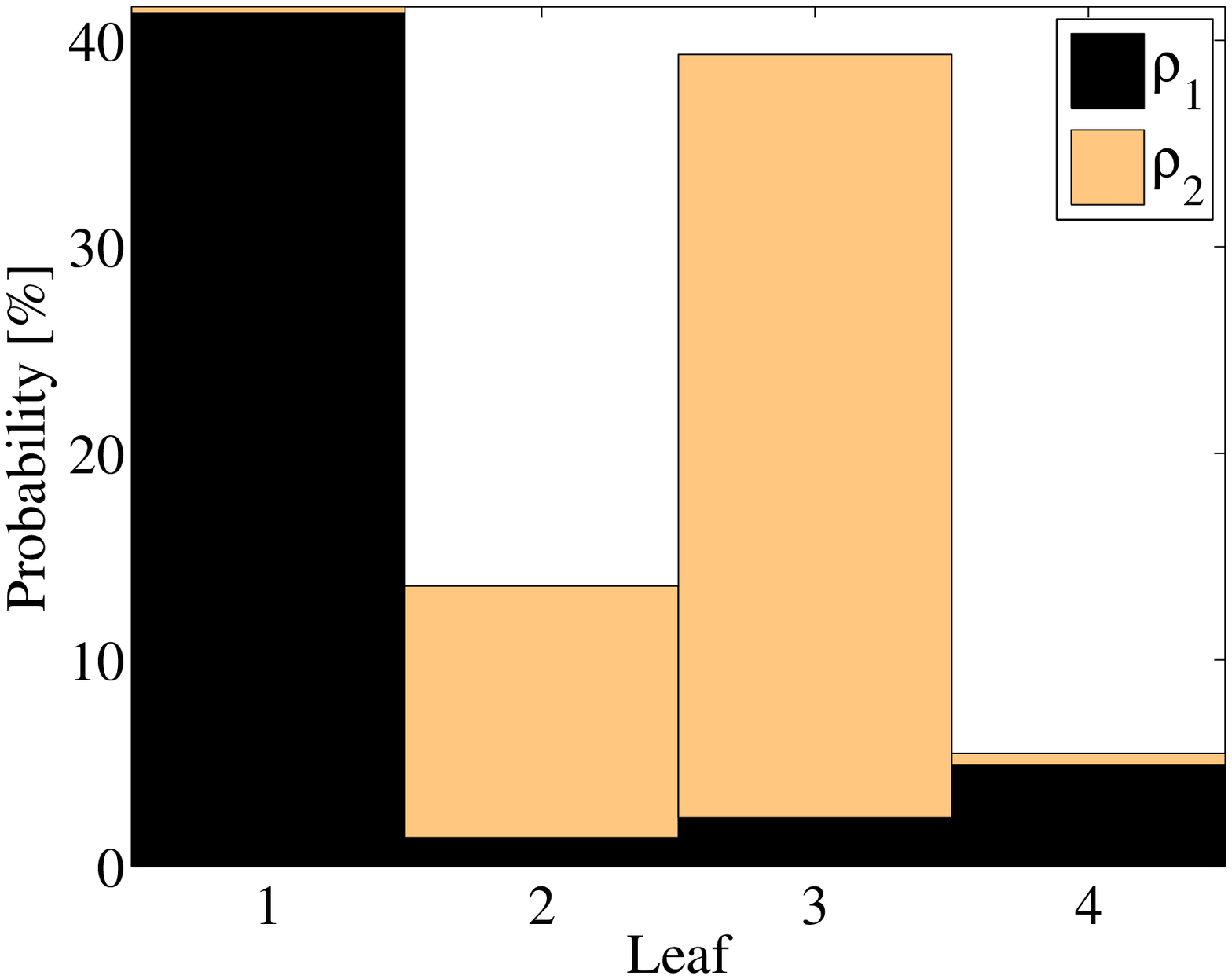} & \includegraphics[width=0.250\textwidth]{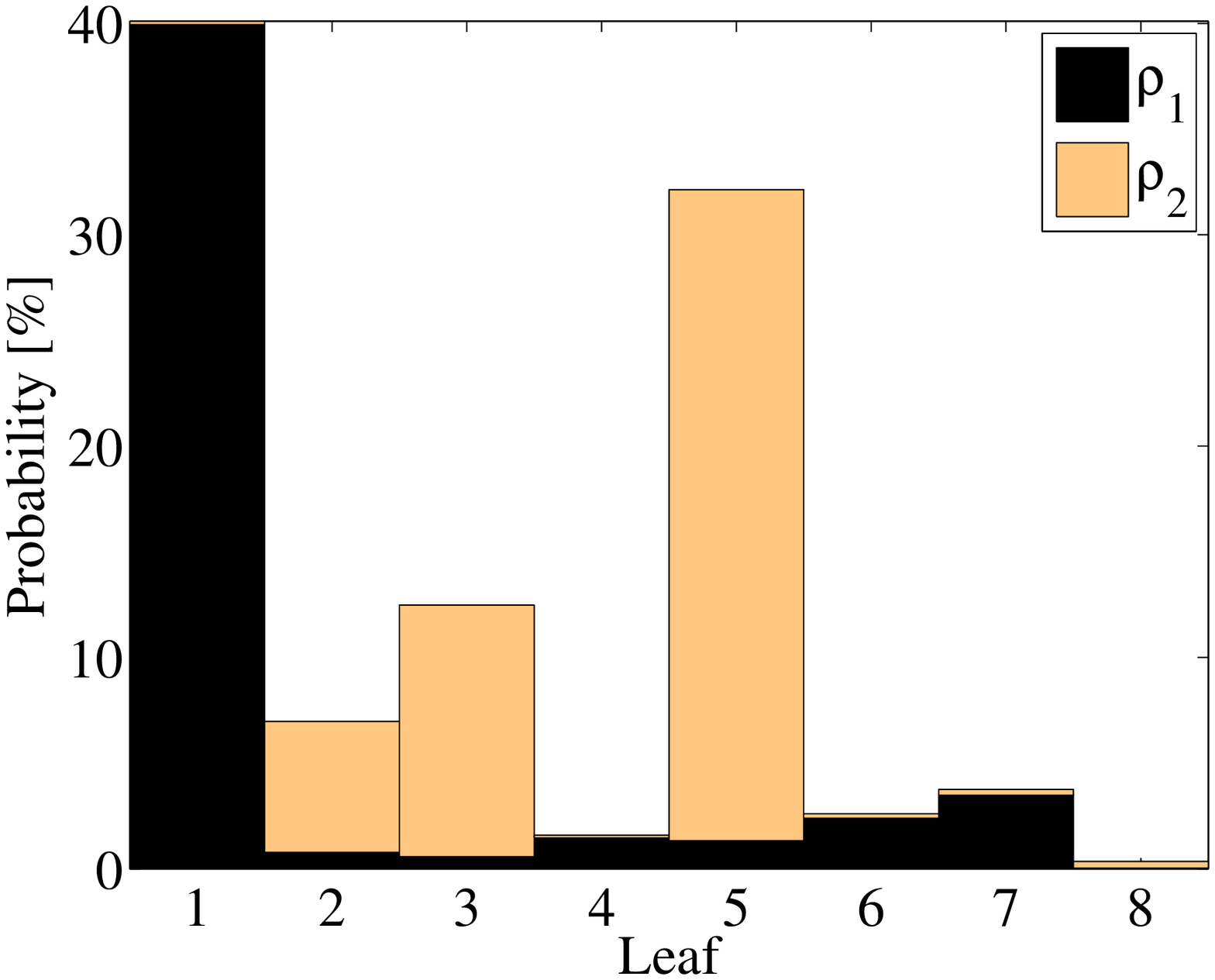} & \includegraphics[width=0.250\textwidth]{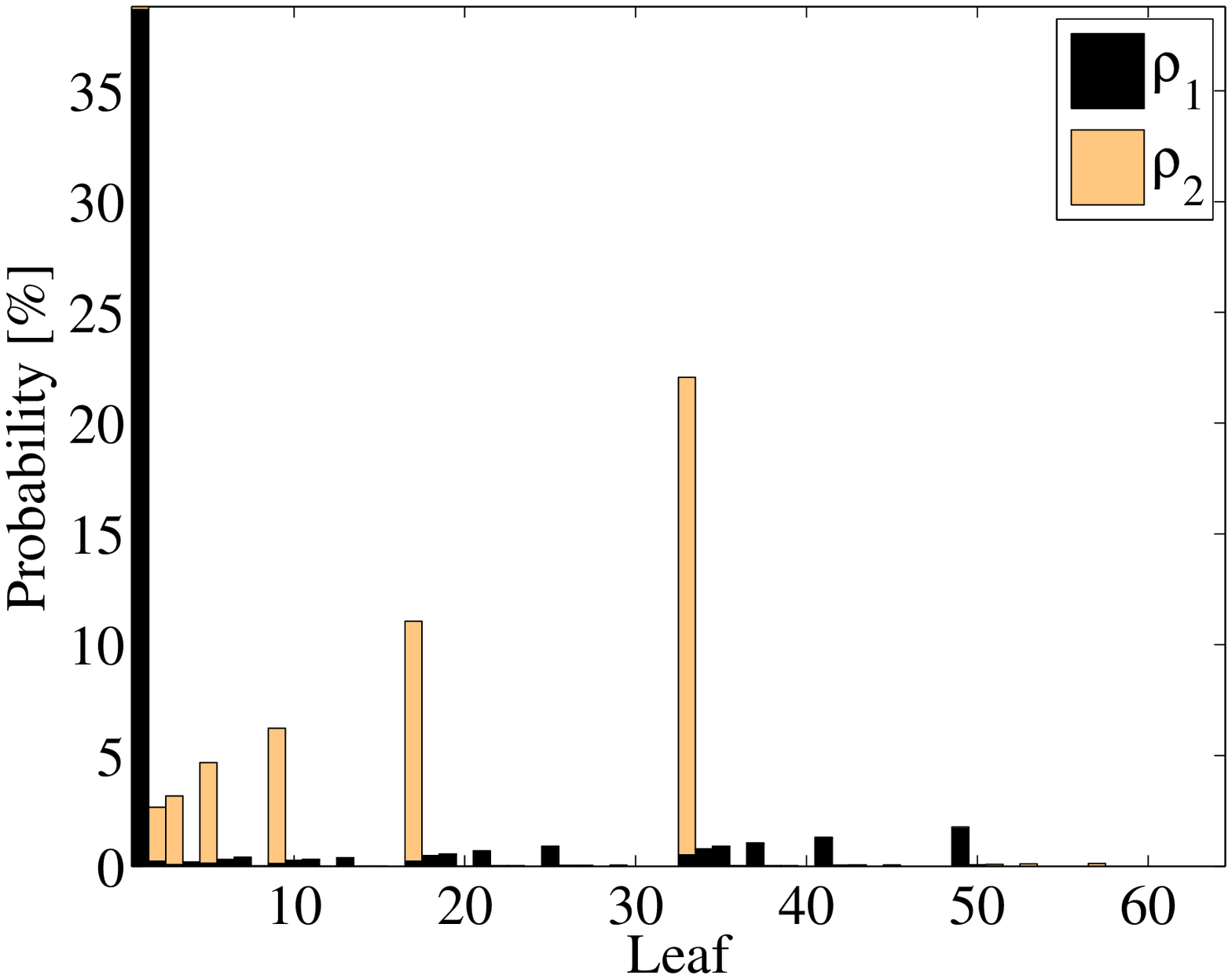} \\ 
{\scriptsize (a) $N = 1$} & {\scriptsize (b) $N = 2$}& {\scriptsize (c) $N = 3$} & {\scriptsize (d) $N = 6$}   
\end{tabular}
\caption{Probability distributions obtained at the leaves of the decision tree from the two candidate qubits of Eq. (\ref{eq:OurQubits}). At each partial measurement, the ancillary modes underwent a coherent displacement and were projected by an APD of unit quantum efficiency. The leftmost histogram (a) is obtained by a direct measurement ($N = 1$) while histograms (b), (c), and (d) are obtained for $N$ equal to 2, 3, and 6, respectively. Although the figures of merit improve with increased $N$ (cf. Fig. \ref{fig:CD_APD}), we see that the larger trees also ``waste'' many computations to branches of little statistical significance. This is particularly salient in (d): Except for sporadic spikes in the probabilities, most branches that slope toward the right provide little insight. To remedy this unnecessary overhead in computational resources, it is better the abort the recursion than aimlessly persevere to lower levels with little or no overall gain in the figures of merit. Note also an interesting structure in the probability distributions with $\rhohat_1$ dominating the no-click sequence of outcomes and $\rhohat_2$  obeying an ostensibly periodic pattern.}
\label{fig:PDF_C_2}
\end{figure*}

Another computational overhead is due to the \textit{fragmentation} of the Bayesian network \cite{Murthy1998}. This is the process whereby vast swaths of the decision tree yield little statistical significance and yet take up as much resources to compute as the more relevant branches. This is illustrated in Fig. \ref{fig:PDF_C_2}. In the case of the top-down greedy algorithm we have implemented, this could have been averted by aborting the recursion as soon as the statistical significance of a given node, $\sum_{c} p_{c}^{(k,\nu)}$, falls below a certain threshold, or alternatively, if the figure of merit fails to converge at a satisfactory rate from one level to the next. Such a resource management technique would result in unbalanced trees, whose leaves may not all lie at the deepest level $k = N$.

%%%%%%%%%%%%%%%%%%%%%%%%%%%%%%%%%%%%%%%%%%%%%%%%%%%%%%%%%%%%%%%%%%%%%
\section{Discussion and outlook}
\label{sec:Outlook}

By the present work, we have tried to bring some clarity and structure to the design and optimization of generalized measurement. We have stated the problem as follows: Given a selection of variable unitary operations and von Neumann projectors, how can can we assemble them so as to optimize certain figures of merit resulting from the measurement? We have built up an algebraic answer to this question in three stages. The first recognizes that the variable unitary operations can be used to emulate, albeit approximately, a similarity transformation between the von Neumann projector and the ideal POVM. I.e., we have reduced the notion of measurement to a diagonalization problem. The second stage extends this idea to higher dimensions as per Neumark's theorem, thereby providing better control over the interaction between the measured states and the  measuring operator. This is what we referred to as adaptation in Hilbert space. The sequential use of weak measurements came into effect in the third and final stage where we dealt with the time-dependence of adaptation. This aspect has been extensively treated under various formulations such as stochastic Schr{\"o}dinger equations \cite{Wiseman1993, Wiseman1996}, quantum filtering equations \cite{Bouten2009}, or Markov filtering equations \cite{Somaraju2013}. We instead opt to for a probabilistic graphical model---i.e., a Bayesian network---to represent the optimization of measurement under a gradual collapse scenario. More specifically, we use a class probability tree whose leaves represent measurement outcomes and are weighted by probability distributions. The various branches $\vecmu$ through which the candidate states ``trickle down'' from the root to the leaves can then be used to retrodict the state identities $c$. Furthermore, the parameter settings  stored in every node of the tree can be looked up by the experimentalist to determine the optimal measurement settings $\vectau_{o}$ as the data acquisition unfolds in real time.

Quantum information holds many promises for the future of computation. At the present time, however, it may rather be classical computer science, and specifically machine learning, which is more likely to advance quantum measurement protocols. This is what transpires from the second part of this article where we have touched on the various algorithms with which the class probability trees are built. For simplicity, we have used the straightforward greedy approach. It is however clear that the most general measurements will require an algorithm design in their own right. Indeed, we have simplified the problem by choosing an equal beam splitting and we left the ancillary modes empty. In addition, the same pairs of unitary operators and von Neumann projectors were recycled in all $N$ stages. Such assumptions for the sake of simplicity need not hold in the general case as we could conceive of a much more elaborate multiplexing of different combinations of unitaries and projectors, as well as asymmetric couplings with complex ancillary states (e.g., squeezed light \cite{Xiao1987}). Even the bundling of the generalized outcomes as in Eqs. (\ref{eq:L1set}) and (\ref{eq:L2set}), or that of the direct outcomes as in Eqs. (\ref{eq:PiHD1}) and (\ref{eq:PiHD2}) can be modified to better serve the figure of merit. In brief, each of these additional degrees of freedom---while leveraging more control over the measurement---introduce an extra layer of complexity in the  optimization algorithms. As for the very construction of the Bayesian network, we have presented followed a top-down flow of the tree structure which replicates   the chronological order of quantum collapses. There remains to investigate whether different graph configurations with, say, a cyclic layout, would present any benefits.

Finally, we have plotted how the figures of merit respond under different configurations and came to the conclusion that, for most of the setups we have tried, generalized measurement offer a distinct advantage over direct measurements. (Pending further investigation, the case where it did not, i.e. homodyning, may simply be due to a poor choice of the quadrature binning or of parameter range, rather than to any shortcoming of the adaptation \textit{per se}.) Overall, these results are particularly promising in light of the fact that a scaling to larger de-localizations $N$ compensates for lower quantum efficiencies. This is a crucial advantage over direct measurements where quantum efficiency is an irremediable hindrance. There remains of course to further analyze the asymptotic behavior with $N$ in order to see if the figures of merit saturate before reaching their theoretical optima.

\begin{comment}
A final word is in order regarding the experimental aspect. We have modeled the main limiting factor with the quantum efficiency $\eta$. A truly realistic feasibility study will however need to incorporate many other factors such as dark counts, as well as timing issues due to the slew rate of the detectors and various delays in the feed-forward process \cite{Berry2001, Berry2001a, Armen2002, Geremia2004, Cook2007}. 
\end{comment}

%%%%%%%%%%%%%%%%%%%%%%%%%%%%%%%%%%%%%%%%%%%%%%%%%%%%%%%%%%%%%%%%%%%%%
\begin{acknowledgments}
A. L. would like to thank Jonas S. Neergaard-Nielsen for valuable pointers on superoperators. 
\end{acknowledgments}

%%%%%%%%%%%%%%%%%%%%%%%%%%%%%%%%%%%%%%%%%%%%%%%%%%%%%%%%%%%%%%%%%%%%%	
\appendix
\section{Superoperators}
\label{sec:AppendixSuperoperators}

The coupling of the zeroth mode with the $k$th mode, followed by the transformation and the collapse of that $k$th mode, transforms the incoming state $\rhohat_{c}^{(k-1)}$ into $\rhohat_{c}^{(k)}$. Instead of having to carry around partial traces, we can represent the whole transformation of the zeroth mode as a superoperator acting on $\rhohat_{c}^{(k-1)}$  (Fig. \ref{fig:RecursionRepresentation}). This transformation is
\beqa
\rhohat_{c}^{(k-1)} & = & \mbox{Tr}_{\mbox{\scriptsize anc}}\Big\{ \Bdag_{t^{(k)}} \tes{\id \otimes \Adag_{\tau^{(k)}}} \tes{\id \otimes \Pihat_{\mu^{(k)}}} \nonumber\\
& & ~~~~~~~~ \tes{\id \otimes \Ahat_{\tau^{(k)}}} \Bhat_{t^{(k)}}\tes{\rhohat_c^{(k)} \otimes \rhohat_{\mbox{\scriptsize anc}}^{(k)}} \Big\}.
\eeqa
If we write
\beqa
\Khat & = & \tes{\id \otimes \sqrt{\Pihat_{\mu^{(k)}}}} \tes{\id \otimes \Ahat_{\tau^{(k)}}} \Bhat_{t^{(k)}} \nonumber\\
& = & \tes{\id \otimes \sqrt{\Pihat_{\mu^{(k)}}} \Ahat_{\tau^{(k)}}} \Bhat_{t^{(k)}},
\eeqa
we get 
\beqa
\rhohat_{c}^{(k-1)} & = & \mbox{Tr}_{\mbox{\scriptsize anc}} \braces{\Khat^{\dagger} \Khat \tes{\rhohat_c^{(k)} \otimes \rhohat_{\mbox{\scriptsize anc}}^{(k)}}} \nonumber\\
& = & \mbox{Tr}_{\mbox{\scriptsize anc}} \braces{\Khat \tes{\rhohat_c^{(k)} \otimes \rhohat_{\mbox{\scriptsize anc}}^{(k)}} \Khat^{\dagger}} \nonumber\\
& = & \sum\limits_{i = 0}^{\infty} \bra{i} \Khat \tes{\rhohat_c^{(k)} \otimes \rhohat_{\mbox{\scriptsize anc}}^{(k)}} \Khat^{\dagger} \ket{i} \nonumber\\
& = & \sum\limits_{i = 0}^{\infty} \bra{i} \Khat \tes{\rhohat_c^{(k)} \otimes \sum\limits_{n,m=0}^{\infty} \bra{n} \rhohat_{\mbox{\scriptsize anc}}^{(k)} \ket{m} \ketbra{n}{m}} \Khat^{\dagger} \ket{i} \nonumber\\
& = & \sum\limits_{i,n,m = 0}^{\infty} \bra{n} \rhohat_{\mbox{\scriptsize anc}}^{(k)} \ket{m} \bra{i} \Khat \, \tes{\rhohat_c^{(k)} \otimes \ketbra{n}{m}} \, \Khat^{\dagger} \ket{i} \nonumber\\
& = & \sum\limits_{i,n,m = 0}^{\infty} \bra{n} \rhohat_{\mbox{\scriptsize anc}}^{(k)} \ket{m} \bra{i} \Khat \ket{n} \, \rhohat_c^{(k)} \, \bra{m} \Khat^{\dagger} \ket{i} \nonumber\\
& = &\sop^{(k)} \hak{\rhohat_c^{(k)}},
\eeqa
where the Dirac bras and kets only apply to the ancillary mode.

%%%%%%%%%%%%%%%%%%%%%%%%%%%%%%%%%%%%%%%%%%%%%%%%%%%%%%%%%%%%%%%%%%%%%	
\bibliography{mybibfile}

\end{document}